\documentclass[graybox,natbib,nosecnum]{svmult}
\bibpunct{(}{)}{;}{a}{}{,} 

\pdfoutput=1   

\usepackage[T1]{fontenc}
\usepackage{amsmath}

\usepackage{mathptmx}       
\usepackage{helvet}         
\usepackage{courier}        
\usepackage{type1cm}        
\usepackage{subfig}
\usepackage{makeidx}         
\usepackage{graphicx}        
\usepackage{multicol}        
\usepackage[bottom]{footmisc}
\usepackage[normalem]{ulem}	
\usepackage{floatrow}
\usepackage{soul}   

\usepackage{txfonts}

\usepackage{IEEEtrantools}

\usepackage{wasysym}
\usepackage{marvosym}

\usepackage{mathtools}

\usepackage{hyperref}             
\hypersetup{pdfauthor=Burn \& Mordasini}
\hypersetup{backref=true, pagebackref=true, hyperindex=true, breaklinks=true,colorlinks=true,urlcolor=blue, linkcolor=blue,  citecolor=blue,pagecolor=red, bookmarks=true, bookmarksopen=true}

\def\mearth{M_\oplus}
\def\rearth{R_\oplus}
\def\msun{M_\odot}

\def\f1{f_{\rm I}}

\def\beq{\begin{equation}}
\def\eeq{\end{equation}}

\def\t2{\tau_{\rm II}}

\def\sigmas0{\Sigma_{\rm s,0}}
\def\mj{M_{\textrm{\tiny \jupiter }}}

\definecolor{snsgreen}{rgb}{0.0, 0.620, 0.451}
\newcommand{\rev}[1]{#1}

\makeindex             

\begin{document}

\title*{Planetary population synthesis}
\author{Remo Burn \& Christoph Mordasini}
\authorrunning{R. Burn \& C. Mordasini}
\institute{Remo Burn \at Max Planck Institute for Astronomy, Heidelberg, K\"onigstuhl 17, 69117 Heidelberg, Germany. \email{burn@mpia.de},\\
Christoph Mordasini \at Physikalisches Institut, University of Bern, Gesellschaftsstrasse~6
CH 3012 Bern, Switzerland. \email{christoph.mordasini@space.unibe.ch}}
%
%
\maketitle

\abstract{The planetary population synthesis method aims at comprehensively testing  planet formation theories against observational evidence and providing theoretical sets of planets to help interpret observations and inform instrument development. Recent developments on the theoretical and observational sides are reviewed: First, observational constraints are summarized, then, the work flow of population synthesis and its two main components are presented, which are, global end-to-end models of planetary formation and evolution and probability distributions for the disk initial conditions. Next, the output of four recent population synthesis models is compared in detail and differences and similarities are discussed. The goal is to help the reader understand the assumptions that were made and how they impact the results. Furthermore, future directions of research are identified and the impact of current and future observational programs is discussed. With JWST, evidence on disk and planet compositions emerges. Planet formation models need to prepare for these near-future developments by including self-consistent magnetic wind-driven gas and dust disk evolution, planetary migration, as well as employ hybrid pebble and planetesimal accretion, which are identified as dominant modes of accretion in different mass regimes.}

\section{Introduction}

In the past few years, since the publication of the previous edition of this handbook \citep{Mordasini2018}, the field of exoplanet sciences has extended the number of known planets; but it also moved to a stage where more details can be learned about individual targets -- it has transitioned from the \textit{discovery} to the \textit{characterization} stage. The NASA TESS mission \citep{Ricker2014} with its {observations} of a large part of the sky identified more than two thousand transiting exoplanet candidates out of which many are suitable to follow-up observations from ground-based facilities. In this way, more planets with both mass (from the ground-based radial velocity follow-up observation) and radius (from the TESS light curve) become known. If photometric follow-up is required, more precise and targeted light curves can be obtained with the CHEOPS spacecraft \citep{Benz2021}. The synergy is exemplified by the recent discoveries of two systems with at least six planets \citep{Leleu2021,Luque2023}.

\rev{We have gained invaluable recent insights about exoplanets and their birthsites, the  {protoplanetary disks} around young stars. Foremost, the Atacama Large (Sub-)Millimeter Array has made possible the resolution of the spacial distribution of dust in protoplanetary disks \citep{ALMA-Partnership2015}. By now, surveys of all close-by star-forming regions have been conducted to provide much clearer constraints on the dust mass in those young objects than what was possible a decade ago \citep{Ansdell2016,Ansdell2017,Williams2019,Tobin2020}. From optical and and near-infrared observations of protoplanetary disks, it was possible to find the first exoplanets which are still embedded in the disk, accrete gas, and are even surrounded by a circumplanetary disk \citep{Keppler2018,Muller2018,Haffert2019,Keppler2019,Benisty2021}.}
	
\rev{With the launch of the James Webb Space Telescope (JWST) it was made possible to use infrared spectroscopy to learn about the gas phase chemical inventory of the innermost few astronomical units in protoplanetary disks where planet formation is potentially ongoing \citep{Perotti2023,Tabone2023,Banzatti2023,vanDishoeck2023,Henning2024}. This will help to constrain the overall evolution of that disks go through \citep{Eistrup2023,Mah2023}. In particular, it requires modeling of compositional evolution. In global models, or models including the final assembly of planets (without evolution) this has been included \citep{cridlandpudritz2016,Cridland2017,Tinetti2018,Pacetti2022,Booth2017,Booth2019,Penzlin2024,Schneider2021} and will be fruitful for the comparison with observations from JWST (see also Chapters by {Bergin, Pudritz} in this volume of the Handbook of Exoplanets). Furthermore, insights on the ice composition can be gained from absorption of stellar light in edge-on disks \citep{Sturm2023,McClure2023}.}

\rev{This does not only apply for the disks, but also for the final planets whose atmospheres can now be probed using transit spectroscopy \citep[e.g.][]{JWSTTransitingExoplanetCommunityEarlyReleaseScienceTeam2023,Alderson2023,Feinstein2023,Ahrer2023,Coulombe2023,Lincowski2023,Madhusudhan2023,Holmberg2024}, that is, measuring the wavelength-dependent transit depth which records imprints of the upper atmosphere composition. With this spectroscopic characterization of atmospheric absorption, a tool to study the inventory of planetary atmospheres has emerged. For giant planets and a small number of the smaller sub-Neptunes, the spectra show detectable features. For rocky planets, atmospheric characterization using transit spectroscopy (also called transmission spectroscopy) with JWST is challenging. Rather, the thermal emission from the potentially rocky surface can give insights on the mineralogy of a small number of hot planets \citep{Greene2023,Zieba2023}.}

\rev{An often mentioned property of the atmospheres is the bulk number ratio of carbon atoms to oxygen atoms (C/O) which would in principle allow to obtain information on the location where the planet accreted most of its envelope. However, the retrieval of these locations is challenging \citep{Mordasini2016,Molliere2022,Bitsch2022,Mordasini2024}. We would like to point out that reports of a bulk C/O measurement obtained from consistent atmospheric modeling facilitates easier comparison as opposed to an estimate using absorption lines originating from different heights and an arbitrary pressure-temperature profile (so-called free retrievals).}

\rev{A more simple quantity, which might be interesting to consider for challenging measurements for smaller planets, is the bulk mean molecular weight of the atmosphere. First results hint at a trend of higher mean molecular weights, and thus higher metallicity envelopes, for lower planetary mass \citep{Kempton2023}. The measurement is model-dependent, but it is well-established that a high mean-molecular weight atmosphere results in a more compact structure and smaller spectral features. Utilizing this principle, \citet{Benneke2024} inferred that approximately half of the envelope mass of the sub-Neptune (2.2\,$\rearth$)-sized planet TOI-270 d consists of heavier species than H$_2$ and He. An extension of this kind of measurement to a statistical sample can inform models of gas accretion, internal structure, and envelope mass loss. The conclusion that the planet must have formed outside of the water iceline -- the location in the disk where ice condenses -- lies close at hand since it could then accrete ices in solid form, which subsequently evaporate and make up a large fraction of the gaseous envelope. However, as emphasized by \citet{Benneke2024}, an alternative explanation could be that the envelope is enriched trough interactions of the atmosphere with the rocky, deep interior \citep[see][for a review]{Lichtenberg2023}.}

\rev{We note that another valuable, fully statistical constraint on exoplanet demographics will become available after the launch of the Nancy Grace Roman space telescope. The survey will probe a so-far inaccessible parameter space of planet occurrence around the water snowline \citep{Penny2019}. In the same direction the census of planets will be extended by the PLATO \citep{Rauer2014} and Earth 2.0 \citep{Zhang2022} missions. In contrast to the distant microlensing planets, those two transit surveys will provide samples of planets suitable for more detailed characterization, which is also the goal of the later ARIEL mission \citep{Tinetti2018}.}

This kind of progress on the observational front has motivated advances in terms of planet formation theory \citep{Raymond2022,Drazkowska2023}, such as the effects of drifting pebble-sized particles, which need to be incorporated in the models used for \textit{Planetary Population Synthesis}. Furthermore, new output, such as both the planetary radius and mass, but also the composition of planets and their atmospheres is required to compare the synthetic populations with the observed ones, which is the key goal of this approach.

This chapter contains a review of recent developments on statistical constraints of exoplanet occurrence and an introduction to the planetary population synthesis approach and its methods. Further, recent results are reviewed and open questions are identified. The aim is to provide a summary of advances, challenges, and limitations rather than a complete description of the modeling work or planet formation \citep[for that, the reader is refered to reviews by][]{Benz2014,Mordasini2018,Raymond2022,Emsenhuber2023a,Drazkowska2023,Weiss2023,Mordasini2024}.
 
\section{Statistical observational constraints}\label{sect:statobsconst}

The {number} and type of  {observational statistical constraints} available for comparisons is in principle very large and multifaceted. \rev{For more in-depth discussion, we refer to Chapter {	Planet Occurrence: Doppler and Transit Surveys} in this volume of the Handbook of Exoplanets or reviews by \citet{udrysantos2007,winnfabrycky2015,Zhu2021,Gaudi2022,Weiss2023,Lissauer2023}}. However, there are a number of key constraints the comparison to which population syntheses have traditionally focused on. 

These key constraints are usually the results of large observational {surveys}, both from the ground and  space. Important surveys and publications analyzing them are, e.g., the HARPS high precision \citep{Mayor2011}, the  Keck \& Lick  \citep{howardmarcy2010}, the California Legacy \citep{Rosenthal2021}, and the CARMENES M dwarf \citep{Sabotta2021} radial velocity surveys; for transits the Kepler \citep{Borucki2010,coughlinmullally2016} and TESS \citep{Ricker2014}  surveys; the various direct imaging surveys (\citealp[reviewed in][]{Bowler2016}, recent additions are the surveys at GEMINI \citealp{Nielsen2019}, SPHERE \citealp{Desidera2021,Vigan2021}, and  NACO-ISPY \citealp{Launhardt2020,Cugno2023})), or the microlensing surveys \citep[][]{cassankubas2012,Suzuki2016}. The high importance of surveys stems from the fact that they have a well-known observational bias. This makes it possible to correct for it and to infer the underlying actual distributions that are predicted by the theoretical models.  

All these different techniques put constraints on different aspects of the global models. Especially when they are combined, they are highly constraining even for global models that often have a significant number of free parameters as the combined data carries so much constraining information. 

The constraints can be grouped into three classes: the frequency of different planet types, the distribution functions of planetary properties, and correlations with stellar properties. Next, a short overview of these observational constraints is given. 

\begin{figure}[h]
    \centering
    \includegraphics[width=\textwidth]{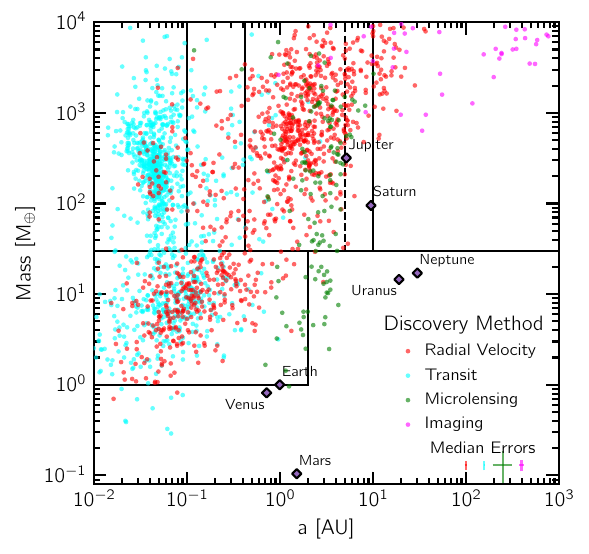}
   \caption{Observed mass-distance diagram of confirmed exoplanets in the NASA Exoplanet Archive as of September 2024. The different colors show the method used for discovery. \rev{Black boxes mark typical definitions for (from top left) hot Jupiters, period-valley giants, warm Jupiters, long period giants, and (bottom) super Earths and (sub-)Neptunes \citep{Chiang2013,Alessi2020}. The dashed line marks 5\,au, which is commonly used to discuss giant planet occurrence studies.} The planets of the Solar System are also shown for comparison\rev{ and median relative semi-major axis and mass errors for each method are indicated at the bottom right.}}\label{CMaMEpoch2017}
\end{figure}

\subsection{Frequencies of planet types}\label{sect:frequenciesobs}
To start, observational constraints on the {frequency of stars with planets} are summarized. In galactic terms, they are available for stars in the Solar neighborhood only (within $\sim$100 pc for radial velocity discoveries and within $\sim$2 kpc for transit ones).
\begin{itemize}
\item The frequency of hot Jupiters \rev{(top left zone marked in Fig. \ref{CMaMEpoch2017})} around solar-like stars is about 0.5-1\% \citep{howardmarcy2010,Mayor2011,Dawson2018}. 
\item The frequency of giant planets within 5-10 AU \rev{(dashed rectangle in Fig. \ref{CMaMEpoch2017})} is 10-20\% for FGK stars \citep{cummingbutler2008,Mayor2011,Fulton2021}. The giant planets have a multiplicity rate of about 50\% \citep{bryanknutson2016}.
\item There is a high frequency (20-50\%) of close-in (fractions of an AU) low-mass (a few Earth masses) respectively small ($R\lesssim4\rearth$) super-Earth and sub-Neptunian planets from high-precision radial velocity \citep{Mayor2011} and the Kepler survey \citep{fressintorres2013a,petigurahoward2013,Zhu2018a,Hsu2019}. These planets are often found in tightly packed multiple systems \citep{Weiss2023}. Planetary systems clearly different from the solar system are thus very frequent.
\item There is a lower frequency on the 1-5\% level of detectable (i.e., sufficiently luminous) massive giant planets at distances of tens to hundreds of AU \citep{Vigan2021}. This means that the frequency of giant planets must drop somewhere with orbital distance by about a factor ten. The occurrence rate is likely positively correlated with the stellar mass \citep{Bowler2016}.
\item There is a high frequency of cold, roughly Neptune-mass planets around M dwarfs as found by microlensing surveys \citep[][]{cassankubas2012,Suzuki2016}.
\item As inferred with the same technique, free-floating or wide-orbit, (super-)Jupiter mass planets are common \citep{Sumi2011,Sumi2023} but it can be assumed that a significant fraction of them were never bound to a star as indicated by the detection of low-mass binary systems using JWST \citep{Pearson2023}, since binarity is not expected for ejected planets.
\item There is a very high total fraction of stars with detectable planets of $\sim$75\% as indicated by high-precision radial velocity searches with a $\sim$ 1 m/s precision \citep{Mayor2011} and almost unity based on the Kepler transit mission \citep{Hsu2019}. At least in the solar neighborhood, stars with planets are thus the rule.
\end{itemize}

\subsection{Distributions of planetary properties}\label{sect:distribobsprop}
\begin{itemize}
\item One of the most important diagrams is the observed distribution of planets in the {mass-distance} (or radius-distance) plane, revealing a number of pile-ups and deserts (see Fig. \ref{CMaMEpoch2017}). For the comparison with the synthetic populations, it is of very high importance to keep in mind that the observed diagram gives a highly distorted impression of the actual population because of the detection biases of the different techniques. Hot Jupiters, for example, appear to be frequent in this plot. But the plot still illustrates the enormous diversity in the outcome of the planet formation process. At the same time, it also indicates that there is structure. 
\item The mass function is approximately flat in log space in the giant planet regime \citep{marcybutler2005} for masses between 30 $\mearth$ and about 4 $\mj$ {(where 1 $\mj$ is the mass of Jupiter)}. At even higher masses, there is a drop in frequency \citep{santosadibekyan2017}. The upper end of the planetary mass function is poorly known, but might lie around 30 $\mj$ \citep{sahlmannsegransan2011}. Towards the lower masses, at around 30 $\mearth$, there is a break in the mass function and a strong increase of the frequency towards smaller masses \citep{Mayor2011}. The mass function below a few Earth masses is currently unknown. 
\item The semi-major axis {distribution} of giant planets consists of a local maximum at a period around 4 days caused by the hot Jupiters, a less populated region further out (the period valley) and finally an upturn at around 1 AU \citep{udrymayor2003,Fulton2021}. The frequency seems to be decreasing beyond 3-10 AU \citep{bryanknutson2016,Fernandes2019,Fulton2021}.
\item The eccentricity distribution is, in contrast to the solar system with its very low eccentricities, broad, including some planets with eccentricities that exceed 0.9. The upper part of the distribution follows approximately a Rayleigh distribution, as expected from gravitational planet-planet interactions \citep{jurictremaine2008}{, indicating together with several other points that in some systems strong dynamical interactions occurred (see the discussion in \citealt{winnfabrycky2015})}. A significant fraction of orbits is however also consistent with being circular. Eccentricities of lower mass planets ($\lesssim30\mearth$) are usually restricted to lower values $\leq 0.5$ \citep{Mayor2011}. Single Kepler planets are on eccentric orbits (mean eccentricity $\bar{e}\approx0.3$), whereas multiples are on nearly circular ($\bar{e}\approx0.04$) and co-planar orbits with mean inclination $\bar{i}\approx1.4$ deg, with $\bar{e}\approx1-2\bar{i}$ \citep{xiedong2016}.
\item The radius distribution of confirmed (Kepler) planets has a local maximum at around 1 Jovian radius as expected from the theoretical mass-radius relation \citep{mordasinialibert2012c}, followed by a distribution that is approximately flat in $\log(R)$ at intermediate radii of 4-10 $\rearth$. Below this radius, there is strong increase in frequency \citep{fressintorres2013a,petigurahoward2013}. At about 1.7 $\rearth$, there is a local minimum in the radius histogram \citep{Fulton2017,Petigura2022} separating super-Earths from sub-Neptunes. This could be due to atmospheric escape of primordial H/He envelopes \citep[][]{owenwu2017,jinmordasini2018} or due to the planets forming with distinct compositions \citep{Venturini2020,Izidoro2022,Burn2024}.
\item While the exoplanet population is characterized by a large diversity between different systems, there is uniformity within individual systems of close-in small planets, known as the peas-in-a-pod pattern \citep{Weiss2018,Millholland2017,Weiss2023}: planets in one system have similar radii, masses, and orbital spacings.
\end{itemize}

\subsection{Correlations with stellar properties}\label{sect:obscorrstellarprops}
\begin{itemize}
\item The best known {correlation of planetary and stellar properties} is the increase of the frequency of giant planets with host star metallicity \citep[][]{gonzalez1997,santosisraelian2004,fischervalenti2005,dongxie2018}. In the super-solar metallicity domain, the frequency of giant planet increases approximately by a factor ten when going from [Fe/H]=0 to [Fe/H]=0.5. This is often taken as indication that core accretion is the dominant mode of giant planet formation \citep{Ida2004,mordasinialibert2012a}. The frequency of low-mass or small planets is in contrast less dependent on metallicity  \citep{Mayor2011,Petigura2022}.
\item The frequency of giant planets is lower for lower mass stars and around 2-6\% for M-dwarfs \citep{bonfilsdelfosse2013,Sabotta2021}. On the other hand, there are 2-3 times more low-mass planets ($\sim$10 $\mearth$) around M-dwarfs than around G-stars \citep{Mulders2015,Sabotta2021}. For stellar masses higher than 1 $\msun$, the frequency of RV-detected giant planets first increases to reach a maximum at around 1.7 to 2 $\msun$, followed by a  drop for even more massive stars \citep{reffertbergmann2015,Wolthoff2022}. 
\item Statistical correlations with stellar age are  not yet well explored, but a number of detections of close-in planets around T-Tauri and young PMS  stars have occurred \citep{mannnewton2016,davidhillenbrand2016,donatimoutou2016,yudonati2017}. They show that close-in massive planets already exist after a few Myr, likely indicating  orbital migration via planet-disk interactions. Hot Jupiters might be more frequent around T Tauri stars than main sequence stars \citep{yudonati2017}. At large orbital distances, direct imaging also probes young planets with ages of a few 10 Myr. The PLATO survey will put statistical constraints on the temporal evolution of the population of transiting planets, adding a new temporal dimension to the constraints. 
\end{itemize}

\section{Brief outline of planet formation processes}

The scope of this subsection includes an introduction to the relevant physical processes and quantities used in modern, global planet formation models. More complete reviews can be found in several chapters of the Handbook of Exoplanets, in \citet{Raymond2022,Drazkowska2023,Emsenhuber2023a,Mordasini2024} or in more technical comprehensive model descriptions \citep{Liu2019,Emsenhuber2020a,Kimura2022a}. For all sub-modules to be usable for the population synthesis approach, they need to be of low enough computational cost to simulate the full time evolution and thousands of systems. A key aspect is, however, that several processes, namely the gas disk evolution, solid and gas accretion, as well as orbital migration of the planets happen on similar timescales. Therefore, the modules need to be evolved at the same time and be self-consistently coupled to each other.

\subsection{Gas disk}
Young stars are surrounded by a protoplanetary disk made from gas and dust. Following the realization that older, Class II objects do not contain enough mass in dust to grow planets \citep{Manara2018}, the starting point of global simulations has shifted to earlier times, typically to the Class 0 or Class I stage of protostar evolution \citep[for a modern definition, see e.g.][]{Williams2019}. At this time, the star and disk are less than 100\,kyr old and to some considerable (Class 0) or minor (Class I) degree surrounded by a more extended, spherical envelope of gas remaining from infall. For rapid processes, such as dust evolution and gravitational instabilities, the growth of the disk over time should be included \citep{Birnstiel2010,Schib2021,Schib2023}.

Since these objects are observable, especially in the millimeter to centimeter wavelengths with ALMA and the VLA, estimates for the initial conditions can be derived from observations and are reviewed in a section dedicated to initial conditions below. The initial disk surface density \rev{$\Sigma_{\rm ini}$} profile is in approximate agreement with observations \citep{Andrews2018} described as a power-law with an inner \rev{$r_{\rm in}$}, and outer \rev{$r_{\rm out}$}, exponential cut-off \rev{
\begin{equation}
\Sigma_{\rm ini} =  \Sigma_{0} \left(1-\sqrt{\frac{r_{\rm in}}{r}}\right) \left(\frac{r}{r_{\rm 0}}\right)^{- \beta} \exp\left[-\left(\frac{r}{r_{\rm out}}\right)^{(2-\beta)}\right]\,,
\end{equation}
with a choice of $\beta$ around 0.9 matching observed disk profiles \citep{Andrews2010}. The radius $r_{0}$ (e.g. 1\,au or 5.2\,au) is a reference radius at which the surface density is equal to the normalization surface density $\Sigma_{0}$. The parameters depend on stellar mass and are for planetary population synthesis chosen to match total gas disk masses inferred from observations (see discussion in the Probability distribution of disk initial conditions Section below).}

Allowing for global planet formation simulations requires modeling the full extent of the disk lifetime which is currently only possible with an analytical or one-dimensional model for disk evolution. If the disk evolves as a {viscous disk}, the relevant diffusion equation for the surface density of gas $\Sigma$ as a function of time $t$ and distance from the star $r$ was derived from fundamental conservation laws of angular momentum  \citep{Weizsacker1948,luest1952,Lynden-Bell1974,Pringle1981}
\beq\label{eq:lyndenbellsigma}
\frac{\partial \Sigma}{\partial t}=\frac{1}{r}\frac{\partial}{\partial r}\left[3 r^{1/2} \frac{\partial}{\partial r}\left(r^{1/2}\nu \Sigma\right)\right]-\dot{\Sigma}_{\rm phot}(r)-\dot{\Sigma}_{\rm planet}(r).
\eeq
with a viscosity $\nu$ that is usually written in the $\alpha$-parametrization as $\nu=\alpha c_{\rm s}(T) H$ with $c_{\rm s}(T)$ the isothermal sound speed for a given disk temperature $T$, $H = c_{\rm s} / \Omega_{\rm K}$ the vertical scale height \citep{Shakura1973}, and $\Omega_{\rm K}$ the Keplerian orbital velocity. \rev{The $\alpha$ viscosity prescription assumes turbulence to exist and its associated average Reynolds stress tensor is still uncertain and can be obtained from numerical simulations \citep{Lesur2023}. The collection of indirect observational evidence of turbulence levels in disks is, since ALMA and JWST, a currently ongoing process \citep{Villenave2022,Dullemond2022,Franceschi2023,Flaherty2024,Paneque-Carreno2024,Duchene2024}. } Eq. \eqref{eq:lyndenbellsigma} in this form assume no self-gravity of the disk. Besides the viscous evolution term, the effects of mass loss by photoevaporation (internally- or externally-driven) \citep[e.g.,][]{Clarke2001,alexanderpascucci2014,Haworth2018,Picogna2019,Picogna2021,Ercolano2021} represented by $\dot{\Sigma}_{\rm phot}(r)$ and of gas accretion by the planets giving raise to the $\dot{\Sigma}_{\rm planet}(r)$ term are also to be included. Alternative, more recently preferred models of low disk viscosity also include an advective term on the right hand side of equation \eqref{eq:lyndenbellsigma} \citep{Weder2023}
\begin{equation}
	\label{eq:magnetic_wind}
    \propto \frac{1}{r}\frac{\partial}{\partial r} \left( r \alpha_{\phi,z} c_{\rm s} \Sigma \right)\,,
\end{equation}
where $\alpha_{\phi,z}$ is \rev{the component of the Maxwell stress tensor} exerting a torque on the \rev{surface of the disk. It should not be confused with the viscous $\alpha$. Similarly to the viscous case, it is found in numerical magneto-hydrodynamical simulations where magnetic winds are launched at this location \citep{Blandford1982,Pudritz1986,Bai2013,Bai2016,Bai2016a}. To include this effect in global, one-dimensional disk models, the term in Eq. \eqref{eq:magnetic_wind} was introduced} \citep{Suzuki2016Winds,Tabone2022a,Alessi2022,Weder2023}. In addition, another mass sink term parametrizing the launch of particles in a magnetic wind -- and thus conserving angular momentum -- needs to be considered. Given little diffusivity and strong advection, these modifications lead to a fundamental change in disk evolution from a diffusive disk spreading in two directions to an advection-driven evolution which moves all gas toward the star.

Equation \eqref{eq:lyndenbellsigma} can be directly solved on a suitable grid for planet formation studies. Alternatively, a steady-state solution can be found by imposing that the diffusive term vanishes, leading to a constant $\nu \Sigma$ in radius. For a viscous disk, conservation equations then imply a radial mass flux, or accretion rate, of $\dot{M} = 3\pi\nu\Sigma$. If the temperature structure \citep{Nakamoto1994,Chiang1997,Dullemond2013} and therefore $\nu$ is known, this so-called steady-state disk allows for an analytic calculation of $\Sigma = \frac{\dot{M}}{3\pi\nu}$ based on the accretion rate. Given that the accretion rate of the star can be inferred from observations \citep{Hartmann2016}, this is a useful treatment. A similar solution has been found for wind-driven disks \citep{Chambers2019}. However, the approach cannot capture variations in initial profiles and the removal of mass by planets or photoevaporation in a consistent manner.

\subsection{Dust, pebbles, and planetesimals}
Interstellar dust \citep{Mathis1977} is inherited and transported to the protoplanetary disk as the star is forming \citep{Drozdovskaya2014,Hartmann2016}. Growth of dust will occur, potentially already during infall but surely after the dust has arrived in the gaseous disk \citep[see][for recent reviews]{Birnstiel2016,Drazkowska2023}. {Dust growth} is estimated, based on laboratory works, to be efficient up to centimeter sizes. The appropriate quantity to describe the dust size is the Stokes number, which describes decoupling from the gas motions with increasing Stokes number: $\rm{St} = \frac{\pi r_{\rm p} \rho_{\rm p}}{2 \Sigma}$, for the Epstein drag regime with dust particle size $r_{\rm p}$ and bulk density $\rho_{\rm p}$. For dust and pebbles, the Epstein regime is most of the time applicable but larger bodies can transition to different drag regimes \citep[e.g. boulder-sized objects, see][]{Burn2019}.

The following stages of dust processing occur: particles with Stokes numbers larger than the dimensionless turbulence parameter $\delta_z \sim \alpha$ in vertical direction \citep{Youdin2007} will not be stirred by turbulent motions of the gas and settle toward the midplane. The enhanced number densities promote further collisions, which will lead to sticking up to sizes at which particles bounce-off of each other \citep{Seizinger2013,Arakawa2023} or velocities at which they break up into fragments. While the former limit is often neglected in global models, the fragmentation threshold is considered in the popular two-population approximation \citep{Birnstiel2011} (small dust and larger pebbles) as a limit to the maximum size ${\rm St}_{\rm max} \simeq \frac{v_{\rm frag}}{\alpha c_{\rm s}}$, where the parameter $v_{\rm frag}$ has to be determined experimentally \citep{blumwurm2008,Musiolik2019,Steinpilz2019}. Typical values for ${\rm St}_{\rm max}$ range from 0.01 to 0.1.

Since those Stokes numbers approach unity, the dust decouples from the gas motion that is sub-Keplerian due to pressure support. Therefore, the particles are subject to aerodynamic drag and spiral -- toward the star \citep{Whipple1972,Weidenschilling1977}. The radial velocity of a particle with Stokes number ${\rm St}$ can be obtained from their equations of motion following \citet{Nakagawa1986}
\begin{equation}
v_{\rm r} = \frac{v_\mathrm{g}}{1+\mathrm{St}^2} - \frac{2\eta v_\mathrm{K}}{\mathrm{St} + (\mathrm{St} \varrho^2)^{-1}}\,,
\label{eq:radial_drift_w_feedback}
\end{equation}
where $v_\mathrm{g}$ is the velocity of the gas obtained from solving Eq. \eqref{eq:lyndenbellsigma}, \begin{equation}
\eta = - \frac{r}{2v_\mathrm{K}^2 \rho_{\rm g,mid}} \frac{\partial P}{\partial r}
\end{equation}
is a dimensionless measure of the radial pressure gradient in the disk, and  $\varrho$ is a factor controlling the behavior for large dust to gas ratios given by \citet{Nakagawa1986} becomes important for large midplane dust to gas ratios.

This process called {\textit{radial drift}} supplies the inner disk with solid particles at those Stokes numbers, which are called {\textit{pebbles}}. Their evolution over time can be tracked with a transport equation, with an approximative treatment to obtain representative sizes for dust and pebbles \citep{Birnstiel2012}, or using a full coagulation model \citep{Stammler2022}. In light of the observability of inner disk compositions with JWST \citep{Perotti2023,Banzatti2023,Gasman2023,Tabone2023,Grant2023}, it is interesting to note that pebble drift also implies that volatile elements are released to the gas phase if pebbles cross the respective icelines \citep{Drazkowska2017,Burn2019,Booth2019,Mah2023}.

While the pebble flux can be accreted directly by growing planets (see following section), they are also expected to collapse to larger \textit{planetesimals}. The process of {planetesimal formation} was long disputed due to the inability to trigger planetesimal formation for standard conditions in a smooth disk. However, several mechanisms can concentrate particles spontaneously which can then lead to the gravitational collapse of a cloud of dust and pebbles into larger planetesimals \citep[see][for a recent review]{Lesur2023}. For global models, this process needs to be parametrized, for example with the approach of \citet{Lenz2019} who convert a fraction of the pebble flux to planetesimals at all locations. More distinct locations are favored by self-consistent global models, such as the location where solid water ice sublimates from pebbles and releases the smaller silicate grains with lower Stokes numbers and therefore inefficient drift. This leads to a traffic-jam and potential site of planetesimal formation \citep{Ida2016,Drazkowska2017,Schoonenberg2017,Hyodo2019,Ida2021}.

Of key importance for subsequent solid accretion is the initial {planetesimal size}. If they form via the collapse of gravitationally bound clumps of solid material (i.e. mainly pebbles) the total mass of the clump equates to planetesimal sizes on the order of about a hundred kilometers at 1\,AU up to a thousand kilometers at larger distances \citep{Johansen2009,Simon2017,Schafer2017,Abod2019,Li2021}. This scaling was adopted for global models \citep{Liu2020a,Coleman2021} of planet formation where they play a key role in bridging regimes with inefficient solid accretion rates. \rev{However, simulations which follow and resolve the solids in an individual bound clump found that mass can be lost from the initial clump size \citep{Nesvorny2021,Polak2023}. Potentially, the collapse can also be triggered as soon as the critical conditions are reached, which also results in a reduction of the mass of the largest planetesimal that forms.} Indeed, \citet{Polak2023} find most mass in hundred kilometer-sized (and not larger) planetesimals at all orbital distances which poses a challenge to solid accretion as described below.

\subsection{Solid accretion}
The classical scenario for growth from planetesimal size to larger bodies is by mutual collisions of planetesimals leading to {planetesimal accretion}. Due to the effect of gravity of the target altering the path of a smaller impactor, that is, gravitational focusing, the most massive body accretes most mass by this process. It is a runaway process \citep{Safronov1969,weidenschillingspaute1997} up to a threshold to where the \textit{protoplanetary embryo}, also called \textit{protoplanet} or \textit{oligarch}, locally perturbs the orbits of the planetesimals \citep{Ida1993,Kokubo1998,Kokubo2000}. While planetesimals alone have  low inclinations, only perturbed by gas turbulence \citep{Ida2008b}, the interaction with the growing planet will excite them out of the orbital plane of the disk and drastically reduce the collision probability. Therefore, this {oligarchic growth} phase is less efficient. The aerodynamic drag of planetesimals caused by the gas in the disk is the process which works against the excitation \citep{Adachi1976}. Because it scales with the ratio of surface area to mass of the planetesimal, it is more efficient for small planetesimals. Since the oligarchic growth phase was discovered, the process of planetesimal accretion with $\sim$ 100 km size was known to be able to grow giant planet cores (i.e. an object of about 10 Earth masses) at 5-10 AU only for planetesimal surface densities significantly enhanced over what was inferred to be required to form the solid cores of the Solar System planets (the so-called Minimum Mass Solar Nebula, \citealp{Weidenschilling1977a,Hayashi1981}). The situation is different if the planetesimals are sufficiently damped by drag, that is, if they are kilometer-sized \citep{Thommes2003,Fortier2007,Fortier2013}. For small planetesimals, or in the inner system \citep{Voelkel2020}, the process is however sufficiently efficient, and continues until the accessible reservoir of planetesimals is depleted at the so-called \textit{planetesimal isolation mass} \citep{Lissauer1993}.

To some degree, the issue is alleviated by drag and fragmentation of the planetesimals in the gaseous envelope of a protoplanet increasing its collisional cross section \citep{Podolak1988,Inaba2003,Podolak2020}. This was taken to the extreme by \citet{Ormel2010} invoking the direct accretion of centimeter sized particles instead. This forms the basis of the {pebble accretion} model, where it should be noted that important aspects thereof (i.e. addressing a similar process of solid accumulation) were already understood earlier \citep{Klahr2006a}. These pioneering works led to a now widely adopted, new model of solid accretion \citep[e.g.][]{lambrechtsjohansen2012,Bitsch2015a,Brugger2018,Ormel2021} where planets can access a larger reservoir of particles, since the pebbles drift on their own towards the planets. Therefore, a planet at a given orbital period will in principle get the access to the full disk of pebbles outside of its orbit.

At low masses, the efficiency of pebble accretion increases with mass of the protoplanet once the accumulation of gas around the planet is of a sufficient size. Drag is then enhanced and particles will settle towards the core \citep{visserormel2016}. While for planetesimals, the inclination is of crucial importance to calculate the collision rate with a protoplanet, for pebbles the scale height of the pebble disk $H_{\rm p}$ determines which portion of the pebble flux the planet can access. Two further intricacies are that not all Stokes number particles are equally accreted \citep{Johansen2017} and that the flux of pebbles is stopped by perturbations in the gas disk since radial drift depends on the pressure gradient in the gas disk. The latter is expected to occur once the planet is massive enough to significantly alter the gas disk profile in its vicinity due to its gravity. At this point, the planet reaches its so-called {\textit{pebble isolation mass}} and will only accrete planetesimals or gas. \rev{For a disk with turbulence $\alpha$, it can be estimated following the two-dimensional simulations of \citet{Ataiee2018} as
\begin{equation}
\label{eq:peb_iso}
M_{\rm iso, peb} = M_{\star} \left(\frac{H}{r}\right)^3 \sqrt{37.3 \alpha + 0.01} \times \left[ 1 + 0.2 \left(\frac{\sqrt{\alpha}}{H/r} \sqrt{\frac{1}{\mathrm{St}^2} + 4}\right)^{0.7} \right]\,.
\end{equation}
Typical values are on the order of ten Earth masses. An expression agreeing on the 20\,\% level \citep[Appendix A in][]{Ataiee2018} was also obtained by \citet{Bitsch2018} using three dimensional simulations and the expression in Eq. \eqref{eq:peb_iso} was generalized to account for eccentric planetary orbits by \citet{Chametla2022}.}

\subsection{Gas accretion}
Protoplanetary gas accretion can be split into two phases, a cooling-limited and a disk-limited phase. If the (total) planet mass is not yet massive, below about 50-100\,M$_{\oplus}$, the planet is called \textit{attached} to the surrounding disk. The outer boundary to its gaseous envelope is given by the disk with modification due to the gas' circulation \citep{ormel2015,Moldenhauer2022,Bailey2023}. During this phase the 1D {internal structure equations} \citep{Bodenheimer1986,alibertmordasini2005} can be solved to find a hydrostatic solution and the gas accretion rate. Apart from the boundary conditions, key ingredients are the sources of energy, or luminosity, for the structure and the energy transport controlled by convection and radiation. During the early stage, the luminosity of the planet is dominated by the solid accretion rate, thus the luminosity is given by the released potential energy per unit time $L_{\rm solid} = G \frac{\dot{M}_{\rm acc,solid} M_{\rm stop}}{R_{\rm stop}}$, where $\dot{M}_{\rm acc,solid}$ is the accretion rate  of solids. Here, the stopping radius $R_{\rm stop}$ and mass encompassed at this radius $M_{\rm stop}$ were introduced to emphasize that the accretion luminosity depends on where accreted particles release their energy. Once gas accretion becomes important, its cooling and contraction also supplies energy to the planet. For the energy budget of the planet, one can then proceed to calculate energy transport which depends in the radiative outer layer of the envelope on the (grain) opacity $\kappa$. This implies that the dynamics of the grains (growth, settling) and their micro-physical structure regulating the opacity are of high importance for the envelope's cooling \citep{Mordasini2014,Ormel2014}. The relevant timescale for this process is the Kelvin-Helmholtz cooling timescale, allowing to write  \citep{ikomanakazawa2000} 
\beq
\dot{M}_{\rm e, KH}=\frac{M_{\rm p}}{\tau_{\rm KH}}
\eeq
where the Kelvin-Helmholtz cooling timescale of the envelope is parametrized as \citep{ikomanakazawa2000} 
\beq\label{tKH}
\tau_{\rm KH}=10^{p_{\rm KH}} {\ \rm yr} \left(\frac{M_{\rm p}}{\mearth}\right)^{q_{\rm KH}}\left(\frac{\kappa}{1 {\rm \ g \ cm}^{-2}}\right)
\eeq
where $p_{\rm KH}$ and $q_{\rm KH}$ are parameters that are obtained by fitting the accretion rate found with internal structure calculations like \citet{Bodenheimer1986,ikomanakazawa2000,mordasiniklahr2014,Emsenhuber2020a}. For example, \citet{Ida2004a} used $p_{\rm KH}=9$ and $q_{\rm KH}=-3$ and neglected the influence of $\kappa$. \citet{mordasiniklahr2014} found $p_{\rm KH}=10.4$, $q_{\rm KH}=-1.5$, and $\kappa=10^{-2}$ g/cm$^{2}$. Once can see from Eq. \ref{tKH} that the accretion rate is a rapidly increasing function of mass, and that gas accretion becomes important once $\tau_{\rm KH}$ becomes comparable to, or shorter than, the disk lifetime, which happens at core masses of about 10 $\mearth$ \rev{\citep[i.e. at the so-called \textit{critical core mass, see}][]{Mizuno1978,Mizuno1980,Pollack1996,Piso2015,Venturini2015}}.

This rapid increase of the gas accretion rate is known as {\textit{runaway gas accretion}}. It gets limited once the surrounding protoplanetary disk can no longer supply gas at the required rate, leading to a \textit{disk-limited gas accretion rate}. At this point, the planetary envelope will contract and is said to \textit{detach} from the protoplanetary disk. The structure changes from approximately spherically symmetric to flattened and the circumplanetary disk forms \citep{ayliffebate2012}. The planetary boundary conditions are then modified as gas will fall onto the planet's surface and get shocked \citep{mordasinialibert2012b,marleau2023}. The detailed criterion for detachment is expressed in terms of a Bondi accretion rate (\citealt{dangelolubow2008,mordasinialibert2012b}) or parameterized from higher-dimensional hydrodynamic simulations of the process \citep[e.g.,][]{Machida2010a,bodenheimerdangelo2013,Choksi2023}.

\subsection{Orbital dynamics and migration}
The gravitational interaction between the gaseous disk and the embedded protoplanets results in the exchange of angular momentum \citep[for recent reviews see][]{baruteaubai2016,Paardekooper2023}, which means that the planets change their semi-major axis, i.e., the undergo {orbital migration} \citep{goldreichtremaine1979,ward1986,linpapaloizou1986}. The angular momentum transfer between disk gas and planets via torques leads in most cases to a loss of angular momentum for the planet which means inward migration. In addition to the classical Lindblad and (non-isothermal) co-rotation torques \citep{goldreichtremaine1979,ward1986,Masset2001,Paardekooper2011,Jimenez2017}, a number of other effects have been identified: Planetesimals-driven migration \citep{levisonthommes2010,ormelida2012}, thermal cold and accretion heating torques \citep{Lega2014,benitez-llambaymasset2015,Masset2017}, torques originating from the dust component in the vicinity of the planet \citep{Benitez-Llambay2018}, and a shift in paradigm for low viscosity disks ($\alpha \lesssim 10^{-4}$) introducing various additional effects \citep{McNally2019}. In this case and at low planetary mass, the co-rotation torque needs to be modified to account for the planet moving and the disk adapting only in a delayed fashion. This gives rise to a dynamical co-rotation torque \citep{Paardekooper2014,pierens2015}. If the planet becomes more massive, vortices and feedback processes become important \citep[see][]{Paardekooper2023}. \rev{For consistency, the shift to lower viscosity should be accompanied by a surface-layer accretion flow due to magnetic winds. In this scenario, works have investigated the migration pattern and found a dependency on the ability of the planet to block the accretion flow \citep{Kimmig2020,Speedie2022,Lega2022}. For magnetic disks, further research was recently motivated by \citet{Aoyama2023} who found local magnetic flux concentrations in the vicinity of a Jupiter-mass planet which alters the gas flow pattern and therefore the torques acting on the planet.}

Classically, in viscous disks at larger planetary masses, the planet carves a gap in its vicinity which leads to a transition from the unperturbed case (Type I) to the so-called Type II migration. For global models, several different descriptions of Type II migration were considered in the literature: {\citet{Ida2004a} consider the angular momentum transfer rate in a viscous accretion disk without planets (the viscous torque or ``couple'' in the terminology of  \citealt{Lynden-Bell1974}) and assume that planets in the type II migration regime act purely as relays that transmit angular momentum also at this rate across their gap via tidal torques. In contrast, the Type II migration description of \citet{alibertmordasini2005} assumes that a planet follows the motion of the gas except for the case that the planet is massive compared to the local disk mass, when the planet is assumed to slow down because of its inertia \citep{alexanderarmitage2009}.
	
More recently, it was proposed that Type II migration proceeds similar to Type I migration but with an overall reduction of the torques based on the depth of the gap (reduction of the surface density) \citep{durmannkley2015,Kanagawa2018}. The shape and depth of the gap can be determined from hydrodynamic simulations \citep{Kanagawa2015,Kanagawa2017}. This prescription was adopted by several global models \citep{Ida2018,Chambers2021,Bitsch2023}. Other global models investigated the influence of dynamical torque \citep{Ndugu2021} and the change of migration in wind-driven disks \citep{Ogihara2018}.

\rev{Another effect exploited by global models is that regions of zero torque are likely to exist in protoplanetary disks. At these locations, contributions from various torques cancel-out to allow for stationary orbits. Since planet migration can proceed rapidly relative to planetary growth, the assumption can be made that planets spend a significant part of their growth phase at persistent planetary traps which are identified to be located at transitions in the radial temperature profile caused by varying the dominant disk heating mechanism or at opacity transitions such as the one expected at the water iceline \citep{Hasegawa2011,Cridland2016}. Complementarily, the effect of local, intermittent sub-structures leading to trapping of migrating planets was investigated by \citet{Coleman2016a} and found beneficial for planet growth.}
	
Lastly, planets can also migrate after the gaseous disk has dissipated via the Kozai mechanism under the influence of an inclined external perturber \citep{Kozai1962,fabryckytremaine2007,Chatterjee2008,Naoz2011}. This provides an alternative formation history for hot-Jupiters to disk migration \citep[][for recent reviews]{Naoz2016,Dawson2018,Fortney2021}.
	
If planets migrate radially at different speeds, they are likely to encounter other planets in the same system. If convergent migration of two planets proceeds on long enough timescales, capture into {mean-motion resonances} is possible and a common outcome in models including {N-body} interactions (\citealp[e.g.][]{pierensnelson2008,Alibert2013}; see also \citealp{Raymond2022} for a detailed review of the dynamical analysis). This implies that mostly the migrating planets in the Earth to super-Earth mass regime get locked into resonant chains as long as the inner edge of the disk can act as migration trap \citep{Ataiee2021}. To explain the observed distribution of period ratios of exoplanets \citep{Fabrycky2014}, it is expected that the chains do not persist. They can break spontaneously or due to the vanishing eccentricity damping from the gas disk as it dissipates \citep{Izidoro2017}.
	
Global models used in population synthesis studies now start to include the full N-body integration explicitly \citep{Alibert2013,Coleman2014,Coleman2016,Lambrechts2019,Emsenhuber2020a} or use analytical prescriptions of statistically likely outcomes of dynamical interactions \citep{Ida2010,idalin2013,Kimura2022a}. The latter approach will result on average, but not for the individual case, in the correct outcome of a close encounter. Therefore, it is suited for the statistical planetary population synthesis approach. Explicitly integrating the orbits of the planets comes at the benefit of also individually resulting in more realistic outcomes, as well as capturing less common cases. However, it comes at a substantial computational cost. Furthermore, the rest of the model usually needs to be extended and adjusted. For example, statistical planetesimal accretion prescriptions need to include a model for competition by several accreting bodies \citep{Emsenhuber2020a}, migration prescriptions need to be validated for eccentric orbits \citep{Coleman2014}, and gas accretion might fundamentally change once an eccentric planet moves with speeds comparable to the sound speed through the disk \citep{Mai2020}.
	
Locking planets in mean-motion resonances is not the only effect of N-body interactions. In the classical scenario for the origin of Earth and the terrestrial planets \citep{Raymond2006,Mezger2020}, as well as in global simulations \citep{Emsenhuber2020b}, the growth of (super-)Earth-mass planets in the inner system goes through a phase of large mutual collisions called {giant impacts}. Although fragments likely split from the bulk of the mass, such as in the scenario for the origin of the Moon, it is often assumed that the bulk of the solid mass remains on the target and perfect merging occurs \citep[see however][]{Cambioni2019,Emsenhuber2020}. The mutual gravitational stirring and giant impacts of planets can in this regime be used to estimate analytically a typical mass scale, the so-called \textit{Goldreich mass}, that planets can reach in the inner system under such conditions \citep{Emsenhuber2023a}. Importantly, it is significantly larger than the classical planetesimal isolation mass and may explain the masses of observed close-in extrasolar super-Earths.
	
Lastly, it is possible that instead of a collision, a close encounter of two planets occurs which leads to scattering to the star, out of the system, or to a different orbit. This is a possible scenario for distant planets via core accretion \citep{Marleau2019} but occurs in relatively few simulated systems \citep{Emsenhuber2020b} and cannot account for multiple giant planets in the same system (and especially not in mean-motion resonance, such as in HR 8799).

\section{Population synthesis method}\label{sect:popsynthmethod}
This section contains a review of the general workflow of the {population synthesis method}, the past development of population syntheses models, the physical processes considered in global formation and evolution models, and finally the probability distributions of the initial conditions.

\subsection{Workflow of the population synthesis method}\label{sect:workflow}
The general workflow of the planetary population synthesis method is shown in Fig. \ref{CMpopsynthworkflow}. There are three main elements: first, and most importantly, the global end-to-end model that predicts observable planetary system properties directly based on parent disk properties. It is based on the result of many different detailed models on individual physical processes like the accretion of solids and gas, orbital migration, the structure of the protoplanetary disk and so on. These detailed models typically study these processes with 2/3D hydrodynamic simulations, contrasting the often low-dimensional approach in the global model. The second element are the Monte Carlo distributions for the initial conditions of the global models that are derived from disk observations, from reconstructions of the disk properties in an equivalent way as done for the minimum mass solar nebula, or from theoretical arguments. Third, tools are needed to apply the observational detection bias and to conduct the statistical comparison with the observed population.  

In general, this comparison will reveal differences between the synthetic and the actual observed population, which are then tracked back to assumptions and results about the governing physical processes as suggested by the individual model, constraining them thereby. The setting of model parameters can also be revised. In case that a synthetic population matches the observations -- at least regarding a certain aspect -- the (underlying) synthetic population can also be used to make predictions about aspects that cannot yet be observed, including the expected yield of future surveys. This serves to design future instrumentation best suitable for constraining formation theory, forming a second feedback loop.

\begin{figure}[tb]
    \centering
    \includegraphics[width=\textwidth]{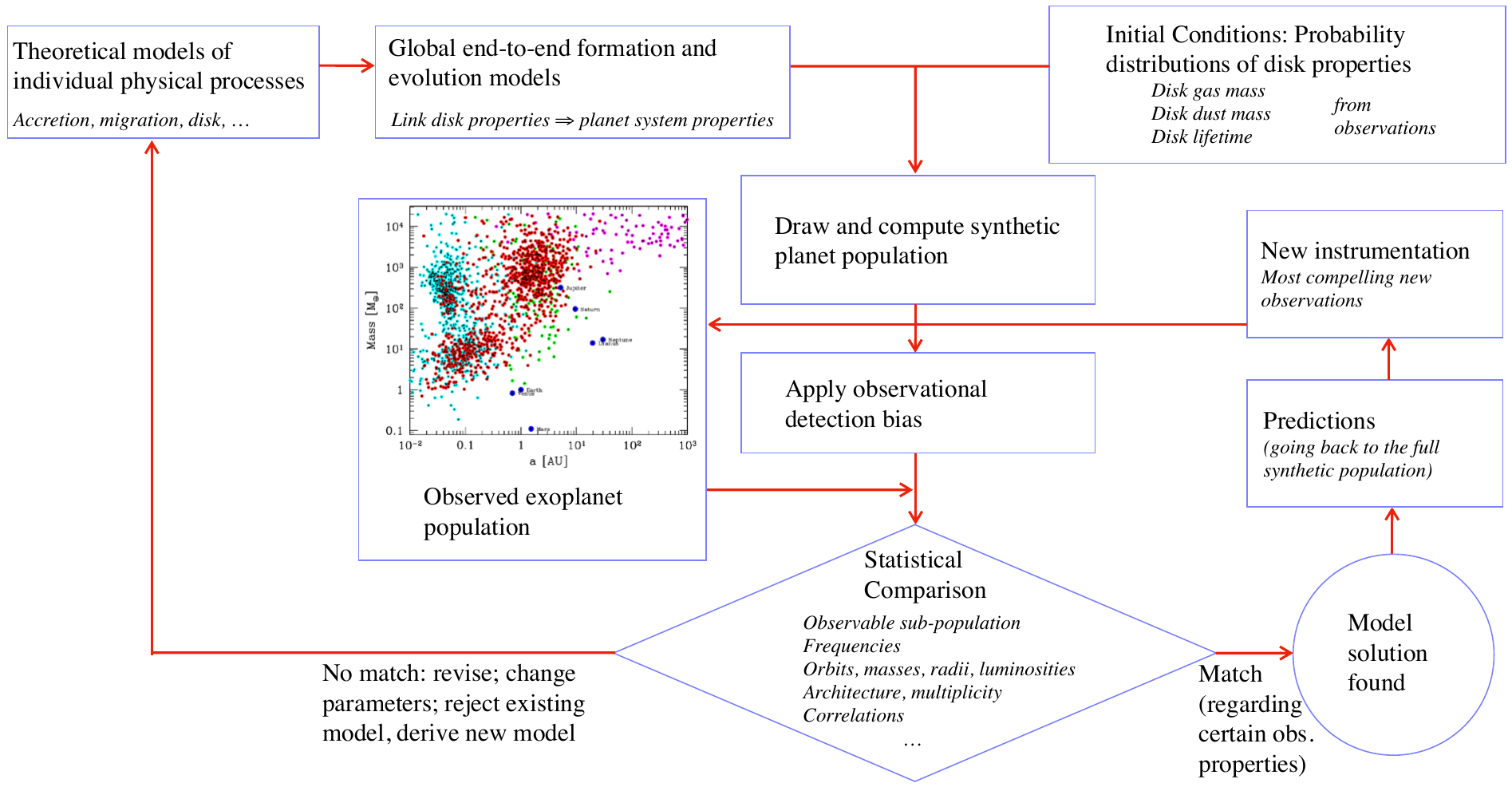}
   \caption{Elements and work flow of a planetary population synthesis framework \citep[updated from][]{Mordasini2015,Emsenhuber2023a}.}\label{CMpopsynthworkflow}
\end{figure}

\subsection{Overview of population synthesis models in the literature}\label{sect:overviewpopmodels}
In other fields of astrophysics, (stellar) population synthesis is a well-established technique for several decades \citep[e.g.,][]{Tinsley1976,bruzualcharlot2003,Vazdekis2016}, while for planets, it is still a relatively {recent} approach. The construction of {planetary population synthesis models} was triggered by the rapidly increasing number of known extrasolar planets. This section contains past and present developments of such models. Early models were all based on the classical core accretion paradigm where the solids are accreted in the form of planetesimals \citep{perricameron1974,Mizuno1980,Bodenheimer1986,Pollack1996}. More recently, models were also based on core accretion with pebble accretion \citep{Ormel2010,Johansen2017}, and on planet formation via gravitational instability \citep{kuiper1951,Cameron1978,Boss1997}.

\begin{enumerate}
\item The Ida \& Lin models. The pioneering population synthesis calculation of \citet{Ida2004,idalin2005a} contained for the first time all the basic elements of population synthesis shown in Fig. \ref{CMpopsynthworkflow}, namely a purpose-built - and therefore fast - global planet formation model based on the core accretion paradigm, and a variation of the initial conditions in a Monte Carlo way. The effects of planetesimal accretion, parameterized gas accretion, and Type II orbital migration in simple power-law disks were considered. As most first-generation population synthesis models the one-embryo-per-disk approximation was used. Later works added Type I migration \citep{Ida2008}, a density enhancement due to a dead zone at the iceline \citep{Ida2008a}, and finally a semi-analytical statistical treatment of the dynamical interactions of several concurrently growing protoplanets \citep{Ida2010,idalin2013}. 
 
\item The Bern Model. Building on the Alibert, Mordasini \& Benz \citep{alibertmordasini2004,alibertmordasini2005} model for giant planet formation in the solar system, \citet{Mordasini2009,Mordasini2009b} presented population syntheses that included quantitative statistical comparisons with observations. Compared to the Ida \& Lin models, the Bern Model explicitly solves the (partial) differential for the structure and evolution of the protoplanetary disk and the planets' interior structure, rather than using power-law solutions. This has the implication of substantially higher computational costs.  Subsequent improvements addressed the structure of the protoplanetary disk \citep{fouchetalibert2012}, the solid accretion rate \citep{Fortier2013}, and the type I migration description \citep{Dittkrist2014}. The model was extended to include the planets' interior structure calculation also in the disk-limited and post-formation phase over Gyr timescales \citep{mordasinialibert2012b}, as well as atmospheric escape \citep{jinmordasini2014}. This makes it possible to predict directly also radii and luminosities instead of masses only. Also these models originally used the one-embryo-per-disk approximation. The concurrent formation of multiple protoplanets interacting via an explicit N-body integrator was added in \citet{Alibert2013}. Recently, the developed modules were fully integrated with a new N-body integrator \citep{Chambers1999} and presented in full detail in \citet{Emsenhuber2020a}. The populations were analyzed in a series of works termed \textit{New Generation Planetary Population Synthesis} (NGPPS) \citep{Emsenhuber2020b,Schlecker2020,Burn2021,Schlecker2021,Mishra2021} and later reviewed \citep{Emsenhuber2023a}.

\item Sharing the disk and gas accretion routines with the Bern Model, but independently developed, \citet{Brugger2018} used pebble accretion as the solid accretion mechanism. This model was also used for a detailed comparison between pebble and planetesimal accretion \citep{Brugger2020} and to explore the hybrid case of combined planetesimal and pebble accretion \citep{Kessler2023}.

\item Recently, another model with planetesimal accretion and resolved envelope structures was presented by \citet{Kimura2022a}. While it shares similarities with the Bern Model, it uses a semi-analytical treatment for resonance trapping and dynamical interactions similar to \citet{Ida2010} instead of direct N-body integration. Furthermore, the study includes envelopes (and their opacities) enriched in water which was outgassed from the planets' magma oceans \citep{Kimura2020}.

\item The models of \citet{hasegawapudritz2011a,hasegawapudritz2012,hasegawapudritz2013}  combine a planet formation model based {initially} on the Ida\& Lin models with power-law disks with inhomogeneities or the analytical disk model of \citet{Chambers2009} {and \citet{cridlandpudritz2016}}. These models emphasize \rev{and utilize for numerical efficiency} the importance of \rev{the zero torque locations (planet migration traps, see Orbital dynamics and migration section), at} the edge of the MRI-dead zone, \rev{the water iceline}, and the transition from the viscously heated to the irradiation-dominated region in the disk. Later updates \citep{alessipudritz2017,Cridland2017,Alessi2020,Alessi2022} include models for dust physics, astrochemistry, magnetic wind driven disks, and radiative transfer. 

\item While not used in population syntheses but in parameter studies, the models of \citet{Hellary2012,Coleman2014,Coleman2016} are global models that combine an N-body integrator with a 1D model for the disk's structure and evolution and the planets' orbital migration. In contrast to other models, the planetesimals are directly included in the N-body as super-particles, and not simply represented as a surface density. Early models use fits to the results of \citet{movshovitzbodenheimer2010} for the planets' gas accretion rate while later models \citep{colemanpapaloizou2017} calculate it by solving 1D structure equations. The model was applied to study the formation of the Trappist-1 system \citep{Coleman2019}, the impact of a planetesimal and embryo formation model \citep{Coleman2021}, and with external photoevaporation in an evolving, simulated cluster \citep{Qiao2023}. Similar global models were also presented by \citet{thommesmatsumura2008a}.

\item Based on the global model of \citet{Bitsch2015a}, \citet{Ndugu2018} presented population syntheses based on the core accretion paradigm where the cores grow by the accretion of pebbles instead of planetesimals. \citet{Ndugu2018} focused on the effect of the stellar cluster environment. The gas disk structure is obtained from 2D simulations including viscous heating and stellar irradiation assuming a radially constant mass flux \citep{Bitsch2015}. The planets' gas accretion rate is given by analytical results of \citet{Piso2014}.  The cores grow by the accretion of mm-cm sized drifting pebbles \citep{lambrechtsjohansen2012,Lambrechts2014a}. The model uses the one-embryo-per-disk approach, such that N-body interactions are neglected, while type I and II migration are included. With a similar model, but using a more simple, steady-state disk, \citet{Liu2019,Liu2020a} conducted studies focusing on brown and M dwarfs \citep[recently also including N-body interactions,][]{Pan2024}, \citet{Nielsen2023} investigated the dependency on the galactic environment, and \citet{Appelgren2020,Appelgren2023,Gurrutxaga2024} revised and studied the impact of the pebble flux model. In \citet{Drazkowska2023}, a new set of populations for Solar mass stars based on this model were presented. The model was also extended to track the composition of planets \citep{Schneider2021,Schneider2022} which was used to study the composition of giant planet atmospheres \citep{Bitsch2022}.

\item The \textsc{PlanetalP} model \citep{Guilera2010,Guilera2014,Guilera2020,Ronco2017} is a modular code used in different settings for parameter studies, but also \citep[e.g. in][]{Ronco2017}, for planetary population synthesis. Dedicated studies using this model were conducted on the effect of planetesimal fragmentation \citep{Guilera2014} and of thermal \citep{Guilera2019,Guilera2021} and dust torques \citep{Guilera2023}. Also, the late evolution of planetary systems was investigated with coupled stellar evolution, tides, and N-body interactions \citep{Ronco2020}. However, gravitational interactions are in most applications of the model not explicitly included, mainly due to computational cost. Gas accretion was both modeled by explicitly coupling the interior structure of the planets -- similar to the Bern model \citep{Venturini2020b} -- or, alternatively, using the prescription of \citet{Ida2004}. A recent advancement of the \textsc{PlanetalP} model was the inclusion of dust growth and transport with consistent pebble accretion of the growing planets applied to planets growing in structured disks \citep{Guilera2020,Sandor2024} and to address the nature of the radius valley using a parametric approach \citep{Venturini2020,Venturini2024}.

\item \citet{Chambers2018} developed a model of planet formation including prescriptions for pebble and gas accretion, migration, and capture into 2:1 mean-motion resonances. The model is optimized towards fast computation time, which nevertheless includes a considerable number of physical processes. This allows him to optimize the parameters of the model towards a better match to observations in an automated way by iterating over $10^8$ model runs, which is a promising pathway to automate the planetary population synthesis exercise.

\item An increasing number of population synthesis calculations are also based on variants of the gravitational instability model for giant planet formation \citep[e.g.,][]{forganrice2013a,nayakshinfletcher2015,muellerhelled2018}. Similar to the core accretion models, these global models couple simple semi-analytic sub-models of disc evolution, disk fragmentation, initial embryo mass, gas accretion and loss (for example by tidal downsizing, \citealt{boleyhayfield2010,Nayakshin2010a}), orbital migration, grain growth, formation of solid cores by sedimentation, and recently, the N-body interaction of several fragments \citep{forganhall2018}.
\end{enumerate}

Table \ref{tab:models} contrasts model parameters of four applications of the models corresponding to items 2, 3, 4, and 7 in the list above. The resulting distributions of planets given the initial condition distributions, also summarized in Table \ref{tab:models}, are then discussed in the Results section below.

\subsection{Probability distribution of disk initial conditions}\label{sect:probdists}

The second central ingredient for a population synthesis calculation are sets of {initial conditions} (see Fig. \ref{CMpopsynthworkflow}). These sets of initial conditions are drawn in a Monte Carlo way from probability distributions. These probability distributions represent the varying properties of protoplanetary disks and are derived as closely as possible from results of disk observations, or, if the quantities are not observable, from theoretical arguments. Typically, there are at least four Monte Carlo variables employed \citep{Ida2004a,Mordasini2009}:

\begin{enumerate}
\item \textit{The metallicity and dust-to-gas ratio} It is  usually assumed that the bulk metallicity is identical in the star and its protoplanetary disk. Then, the disk metallicity [M/H] can be modeled as a normal distribution as observed spectroscopically in the photosphere of solar-like stars in the solar neighborhood, with $\mu=-0.02$ and $\sigma$=0.22 \citep{santosisraelian2005}.  The [M/H] is converted into a disk dust-to-gas ratio
\begin{equation}f_{\rm dg}=f_{\rm dg,\odot} 10^{\mathrm{[M/H]}},\end{equation} with a solar $f_{\rm dg,\odot}$ of about 0.01 to 0.02 \citep{Lodders2003}. Together with the initial disk gas mass and the locations of icelines, $f_{\rm dg}$ sets the amount of solids available in the disk for planet formation. Different assumptions were made on the exact form of the solids. For models using planetesimal accretion (e.g. NGPPS, \citealp{Kimura2022a}, \rev{ \citealp{Alessi2018}}), a large fraction or the full solid content is assumed to be in the form of planetesimals. For pebble-based models, a fraction of the solids is assumed to be in the form of pebbles instead \citep{Brugger2020,Bitsch2015a}. The mentioned planetesimal-based models further assume a 1\% dust-to-gas ratio used for the dust opacity of the disk whereas the pebble-based models more consistently split up the total metal content into dust as opacity source and drifting pebbles.
\item \textit{The initial disk gas mass} The concept of an ``initial'' disk mass is of course questionable as it results from the dynamical collapse of a molecular cloud core \citep{shu1977,huesoguillot2005,Schib2021}, but it could be associated with the disk's mass at the moment when the main infall phase has ended, and no self-gravitational instabilities occur any more. Stability arguments \citep{shutremaine1990}, the inferred mass of the MMSN \citep{Weidenschilling1977a,Hayashi1981}, and observations of protoplanetary disk \citep{Andrews2010,manararosotti2016,Manara2023} point towards disk masses of about 0.1 to 10\% of the star's mass. In recent years, the point in time at which the initial stage is considered have shifted to the earlier Class I or even Class 0 stages as it became clear that the older Class II objects do not contain enough solid mass to grow observed planetary systems \citep{Manara2018}. The radio observations of young objects with significant emission from the more spherical envelope (Class I objects) in long wavelengths using the Very Large Array are particularly useful as they probe deeper into the otherwise (i.e. with ALMA) optically thick emission \citep{Tobin2020,Tychoniec2020}. However, there are large differences between different star forming regions which need to be understood \citep{Williams2019} and neither of the easily accessible nearby clusters are representative of typical star formation in the galaxy which occurs predominantly in denser regions \citep{Winter2022a}.
\item \textit{The disk lifetime} The observations of IR and UV excesses of young stars indicate that the fraction of stars with protoplanetary disks decreases on a timescale of 1-10 Myr, with a classical mean lifetime of about 3 Myr \citep{haischlada2001,mamajek2009}. In a global model, this timescale can either be set directly by modifying the steady-state accretion flow, or is used to find a distribution of external photoevaporation rates that lead together with viscous accretion to a distribution of lifetimes of the synthetic disks that agrees with observations. Recently, the difference between disks in less dense and well characterized environments has shifted to larger values \citep{Michel2021,Pfalzner2022}. As for the initial disk mass, this remains to be investigated in more detail for the interesting denser, but also more distant, star forming regions.
\item \textit{The initial starting coordinates of the embryos} Based on the finding of N-body simulations that oligarchs emerge with relative spacings of a few Hill spheres \citep{Kokubo2000}, a distribution of the starting embryos that is uniform in the $\log$ of the semi-major axis is usually used. It is also possible to arrange the embryos such that they ``fill'' the disk taking into account the asymptotic planetesimal isolation mass \citep{Ida2010}. {In the trapped evolution models of \citet{hasegawapudritz2011a,cridlandpudritz2016,Alessi2018,Alessi2020,Alessi2022} embryos rapidly move into traps, so that it is the locus and movement of the traps that effectively gives the formation locations.} In \citet{Voelkel2021}, a planetesimal formation model \citep{Lenz2019} was consistently linked to the emergence of the embryo. The results highlight prominent formation locations, such as the water iceline, and a significant delay between the bulk of the pebble drift and planetesimal emergence. This implies that the assembly of large enough embryos to start pebble accretion \citep{Voelkel2021a} is often late with respect to the pebble drift. Furthermore, no embryos emerged in the outer disk which poses an additional challenge. Future statistical works, especially when considering the time-sensitive pebble accretion, should include such a delay which is, so far, often parametrized \citep{Drazkowska2023}.
\end{enumerate}

\begin{figure}
	\centering
	\includegraphics[width=\textwidth]{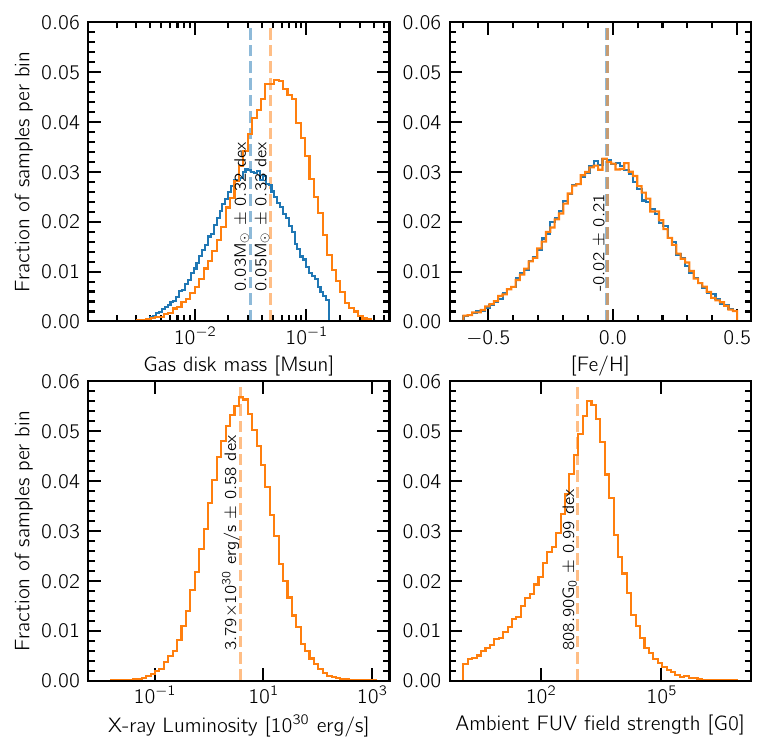}
	\caption{Distributions of initial conditions for disks around 1\,M$_{\oplus}$ stars. \rev{The histograms show for each horizontal bin the number of samples drawn from the probability distributions relative to the total number of draws.} The distributions used in NGPPS \citep{Emsenhuber2020b} which are similar to \citet{Kimura2022a} and \citet{Brugger2020} are shown in blue, suggested distributions by \citet{Emsenhuber2023} are shown in orange. The top two panels show the gas disk mass and the disk metallicity which are parameters in NGPPS. The parameters for realistic photoevaporation models are shown at the bottom and are suggested for future works (see text). Logarithmic mean values are marked and given as text with standard deviations in logarithmic space. Note that the gas disk mass and ambient FUV field strength are not Gaussian distributions.}
	\label{fig:ini_cond}
\end{figure}

Other quantities that may also be treated as Monte Carlo variables are, for example, the quantities describing the initial radial distribution of the gas and solids. For example, the radii of disks seem to be linked to the disk mass \citep{Andrews2010,Andrews2018}, although with a large scatter. Furthermore, this is derived for Class II objects but the initial masses should be taken at earlier times. Therefore, a switch to radii at the same stage \citep{Tobin2020} or to some theory-informed radius \citep{Schib2021,Emsenhuber2023} would make sense. The latter works gauged initial conditions that reproduce observations at later times.

An issue which is often not considered when comparing disk profiles and lifetimes is that in dense environments, disks can be replenished by accreting fresh material from the environment \citep{Kuffmeier2023}. This {environmental effect} is expected based on numerical simulations \citep{Kuffmeier2017,Kuznetsova2020}. This will introduce structures \citep{Kuznetsova2022}, change the mass budget, but also rejuvenate the disk as an older disk can obtain a millimeter spectrum (and thus spectral slope used to classify disks) mimicking a younger object.

In addition, important parameters of the global models are the stellar mass, the planetesimals size, or -- for viscous accretion disks -- the $\alpha$ viscosity parameter \citep{Shakura1973}. They are usually kept constant for one synthetic population, but are varied across different populations to understand their statistical impact in parameter studies \citep[e.g.,][]{Mordasini2009b,Burn2021}. \rev{In \citet{Alessi2022}, values of turbulent $\alpha$ varying from one disk to another was an important element which helped to match observations.}

A change in stellar mass should also be associated with distributional changes of the other Monte Carlo variables. In \citet{Burn2021}, the argument was made for a linear disk mass {scaling with stellar mass} based on extrapolations of trends in disk mass with stellar mass and age \citep{Pascucci2016,Ansdell2017}. A serious issue arises for low-mass disks if external photoevaporation rates are kept constant with stellar mass as can be expected \citep{Haworth2016a,Haworth2018,Winter2022a}. Then, low-mass disks dissipate more rapidly, but a shorter lifetime is not observed \citep{Richert2018}. Therefore, \citet{Emsenhuber2023} concluded that external photoevaporation rates should be reduced to values far below the prediction of the FRIED grid used in their analysis \citep{Haworth2018}. The parameter controlling the external photoevaporation rate is the far ultra violet field strength given in terms of the interstellar value $G_0$.

For internal photoevaporation, if dominated by heating in the X-ray wavelength, the relevant realistic parameter is the X-ray luminosity $L_{\rm X}$, which can be observed \citep{Preibisch2005,Gudel2007} and used in prescriptions of X-ray dominated internal photoevaporation \citep{Picogna2019,Picogna2021,Ercolano2021}.  $L_{\rm X}$ is also an important parameter for the long-term evolution of close-in planets due to photoevaporation of the planetary atmospheres. The stellar mass and time dependency of $L_{\rm X}$ is relatively well constrained by these observations and theoretical works \citep[e.g.][]{Johnstone2021}.

Figure \ref{fig:ini_cond} shows the distributions of initial conditions in the disk which are used or suggested for use in population syntheses. The disk masses shown in blue were found for the Perseus cluster by \citet{Tychoniec2018} and adopted by \citet{Emsenhuber2020a}, \citet{Brugger2020}, and \citet{Kimura2022a} (three of the four models discussed below). Similarly, the disk (and stellar) metallicities based on \citet{santosisraelian2005} were used by the same work to obtain the solid mass accreteable by the planets in the disk. In those works, the external disk photoevaporation was a free parameter tuned to obtain approximately the observed disk lifetime distribution (determined using the criterion by \citet{Kimura2016a}, i.e. when no optically thick region with temperatures above 300\,K remains). The internal photoevaporation was usually not varied among simulations. However, models of disk photoevaporation have made significant advancements and are ready to be used in population synthesis works \citep{Ercolano2022}. Depending on the rotation of the star, the stellar high-energy emission varies which changes $L_{\rm X}$ or $L_{\rm FUV}$ and thus the internal photoevaporation rate. Furthermore, the external photoevaporation was found to occur mainly on the outer edge of the disk instead of an extended flat profile and tabulated rates were published \citep{Haworth2018}. The rate depends on the environment. Classical values for the FUV field strength throughout the galaxy were calculated by \citet{adams2006} and shown in the bottom right panel of Fig. \ref{fig:ini_cond}. It is noteworthy that the distribution is not Gaussian in logarithmic space and that the values are larger due to the larger number of O and B stars than the typical environment in nearby low-mass star forming regions, such as the well-studied Lupus and Taurus clusters.

For this reason, it is unsurprising that \citet{Emsenhuber2023} concluded that a large FUV field strength is inconsistent with observed Class II disk emissions when modeling the time evolution of disks with dust growth and transport as well as entrainment in photoevaporative winds \citep{Sellek2020a,Burn2022}. However, it is not yet clear whether external photoevaporation is not overestimated by the FRIED grid. The aforementioned trend with stellar mass points to a more fundamental issue, such as extinction of the external radiation. If that is the case, the fact that the disk is initially embedded in an envelope and therefore shielded from radiation helps to mitigate this problem and needs to be considered in future works \citep{Qiao2022}.

\citet{Emsenhuber2023} also discuss that an increase in initial disk mass is required to reproduce observed Class II disk emissions. The increase is motivated by the fact that the stellar masses of the Class I sources observed by \citet{Tychoniec2018} are not known. Previous works assumed a Solar mass central star, but this is unlikely since also young sources should follow the initial mass function \citep{chabrier2003}. Therefore, a reduced average mass of 0.3\,M$_\odot$ is well motivated and leads to an increase in disk mass by a factor of three if the disk mass and stellar mass scale linearly. The resulting distribution is shown as orange histogram in Fig. \ref{fig:ini_cond}. It should be noted that with this increase the distribution is more affected by the approximate gravitational instability criterion (a cut at 0.16\,M$_\odot$) which is imposed here and in previous works.

\section{Results: Comparative analysis of population synthesis models}\label{sect:results}
Here, a subset of published results from different research groups are compared. All of those were calculated after the previous version of this review chapter \citep{Mordasini2018}. \rev{Particular emphasis is placed here on topics which were previously inaccessible for population synthesis works: the effects of pebble accretion and dynamical planet-planet interactions.} The selection of models reflects a considerable portion of active groups working on the topic but is not complete. Two models consider that planets grow by pebble accretion, namely the model by \citet{Brugger2018}, which was used for a comparison to planetesimal accretion in \citet{Brugger2020} (B20), as well as the model of \citet{Bitsch2015a} (D23) \citep[specifically, in the form presented in][]{Drazkowska2023}. The other two models use planetesimal accretion, one based on the Bern model of \citet{Emsenhuber2020a} with the Solar mass case discussed in \citet{Emsenhuber2020b} from the New Generation Planetary Population Synthesis (NGPPS) series, and the recent population from \citet{Kimura2022a} (KI22). The latter work focused on M dwarfs in their analysis but here their Solar mass case is shown, which they computed in the course of their project.

\subsection{Formation tracks}
\begin{figure}
	\centering
	\includegraphics[width=1\linewidth]{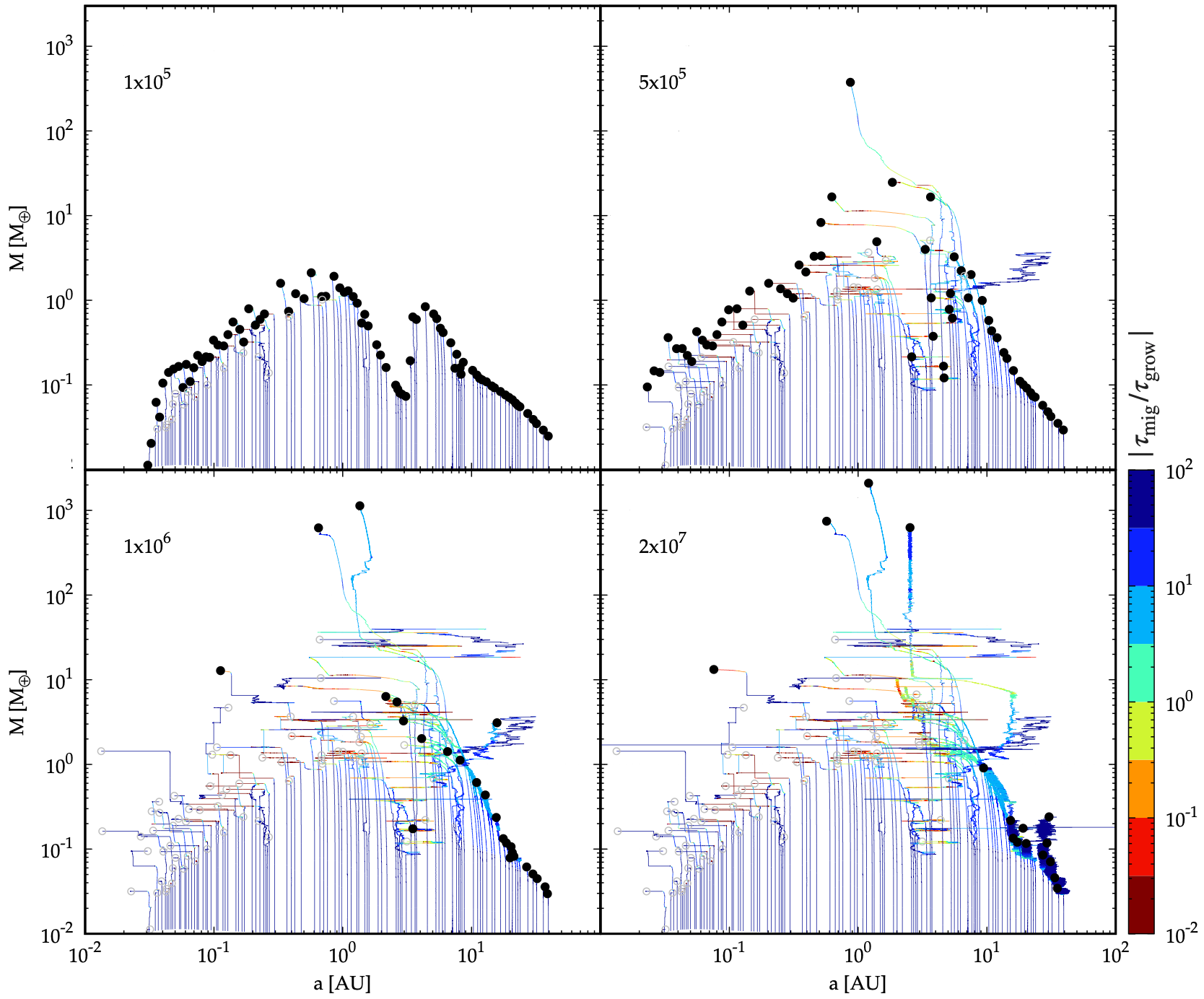}
	\caption{Formation of a synthetic planetary system in semi-major axis versus mass space using the Bern model. The diagram is adapted from \citet{Emsenhuber2020a}. Four time snapshots are shown with full black circles marking the planets, open circles mark the last position of ejected or accreted bodies, and the lines follow their time evolution. The line color scales with the ratio of migration $\tau_{\rm mig}$ to accretion timescale $\tau_{\rm grow}$. In this system, the accretion of large quantities of gas of two bodies emerging close to the water iceline led to a typical destabilization of the inner system with a third body entering the stage of gas accretion afterwards. }
	\label{fig:amtracks852}
\end{figure}
Before the different models are discussed, it is illustrative to visualize the {emergence of planetary systems over time} as a function of the planetary orbital distance and the planet mass. There are differences between the various models but the general patterns are the same. Fig. \ref{fig:amtracks852} shows an example system from \citet{Emsenhuber2020a}. To resolve the history, several snapshots at different times are shown. Here, different stages of planetary system growth are briefly described. For a more detailed discussion, the reader is referred to \citet{Emsenhuber2023a} who reviewed the process in detail and obtained analytically typical relevant mass scales \citep[see also][]{Weiss2023}.

Initially, after placement of the embryos, they grow without significant migration. This is because they are too massive, respectively their surface to mass ratio is too small, to be significantly affected by aerodynamic drag. Furthermore, they are also not yet subject to torques from the gravitational wake, as the gravitational potential of the embryo is negligible up to masses similar to Earth. During this stage, the growth tracks are therefore vertical. As a function of time, there is initially a faster regime during runaway planetesimal accretion before the protoplanet starts to perturb the orbits of the smaller planetesimals in its vicinity and therefore transitions to oligarchic growth. Planetesimal accretion occurs on shorter timescales at short orbital periods, thus the growth in a system generally proceeds inside-out. However, planetesimal accretion without migration stops (reaches \textit{isolation}) when the local reservoir of planetesimals is depleted. This is the case within 1\,AU in the first panel of Fig. \ref{fig:amtracks852} after 100\,kyr. Further out, planetesimal accretion is still proceeding at this time. A prominent feature is the water snowline at 3\,AU where in the Bern model, the bulk density of planetesimals changes by a factor of three and the surface density increases by a factor of two. Therefore, planetesimals are more damped by gas drag and easier to accrete by a growing protoplanet exterior of the water snowline. These factors lead to a significant increase in both the accretion rate and the planetesimal isolation mass.

When embryos, now typically called planets, reach masses on the order of an Earth mass, they start migrating. Due to this, the spacing between embryos, which was initialized to be wider than the range on which interactions occur, decreases. A stage with resonance locking and mutual collisions between embryos occurs. Planets can also be scattered and accreted by the central stars or be brought to large orbits. The typical mass scale of the planets now diverges from the isolation mass in the inner system as discussed by \citet{Emsenhuber2023a}. In contrast, in single planet simulations, this stage can not be recovered and differs significantly \citep[see e.g.][]{Emsenhuber2020a}.

However, the emergence of giants can be reasonably well understood using single planet per disk models. It occurs if the planet reaches large enough masses to trigger runaway gas accretion and almost simultaneously transitions to the slower type II regime. This phase occurs on much shorter timescales and should be considered as a collapse phase. After detaching from the disk and opening a gap, the planet will grow on longer timescales by accreting gas limited by the adopted maximum gas accretion rate criterion (see Table \ref{tab:models}). Due to its mass reaching one or two orders of magnitude larger masses, the emergence of a gaseous giant planet commonly affects the rest of the system by dynamical interactions.

Later evolution of the system mainly proceeds in the outer disk where planetesimal accretion has not stopped. This can lead to the growth of several giant planets and potentially dynamically destabilize the system (as seen in the lower panels in Fig. \ref{fig:amtracks852}). After the gas disk dissipates, systems can also be destabilized, although this occurs more frequently if no giant planet has emerged while the chain of inner planets in resonance before this in systems with a giant planet. Long-term evolution after the gas disk can affect the orbits of the innermost planets by stellar tides, significantly reduces the radii of gaseous planets by cooling and evaporation but usually insignificantly affects their masses. Furthermore, some growth by mutual collisions can proceed in the outer disk over 100\,Myr timescales.

For pebble accretion models, the general picture is identical up to the initial phase. The initial growth generally occurs on a faster timescale and the shape of the typical mass scale after solid accretion is flatter as a function of distance. The innermost planets would also not necessarily be much less massive as the pebble isolation mass is larger than the local Hill-radius limited planetesimal isolation mass. However, pebble accretion inside of the water snowline is sensitive to the size of the ice-stripped silicate grains which is likely small due to their brittle nature \citep{Blum2018}.

\subsection{Initial conditions and parameters}\label{sect:initialconditionssynthesis}

\begin{table}
	\caption{Key model aspects and initial conditions for the populations included in this comparison.}
	\label{tab:models}   
	\begin{tabular}{p{\linewidth/5}p{\linewidth/5}p{\linewidth/5}p{\linewidth/5}p{\linewidth/5}l}
		\svhline\noalign{\smallskip}
		\raggedright \textbf{Model}& \raggedright  \citet{Emsenhuber2020a} Bern Model (NGPPS)& \raggedright \citet{Brugger2018,Brugger2020} (B20) &   \raggedright \citet{Kimura2022a} (KI22) & \raggedright \citet{Bitsch2015a,Drazkowska2023} (D23)&\\
		\noalign{\smallskip}\svhline\noalign{\smallskip}
		\raggedright Solid accretion mechanism &  Planetesimals ($R=300$\,m) & \raggedright Pebble, drift-limited ($0.01 \lesssim \mathrm{St} \lesssim 0.1$), $\alpha_{\rm vert}$:$2\times 10^{-3}$ & \raggedright Planetesimals ($R=20$\,km) & Pebble, $\mathrm{St}$: 0.01, $\alpha_{\rm vert}$:\,$10^{-4}$ & \\
		\noalign{\smallskip}\hline\noalign{\smallskip}
		N-body effects & explicit integration, 100 bodies & None & analytic, 50 bodies & None &\\
		\noalign{\smallskip}\hline\noalign{\smallskip}
		\raggedright Gas accretion method, equation of state, and envelope opacity  & \raggedright Structure Equations, \citet{Saumon1995}, $f_{\rm opa} = 3\times10^{-3}$\,$^\dagger$ & \raggedright Structure Equations, \citet{Saumon1995}, $f_{\rm opa} = 3\times10^{-3}$\,$^\dagger$ & \raggedright Structure Equations, \citet{Saumon1995}, dust profile \citep{Ormel2014} & analytic from \citet{Piso2014}, $\kappa_{\rm env}$:\,0.05\,cm$^{2}$\,g$^{-1}$\\
		\noalign{\smallskip}\hline\noalign{\smallskip}
		\raggedright Gas accretion limit &  Bondi \& local & \raggedright \citet{Machida2010a}\,$\times 0.8$ & \raggedright gap \& non-eq. flow & \citet{Machida2010a} \& flow \\
		\noalign{\smallskip}\hline\noalign{\smallskip}
		Viscous $\alpha$ & $2 \times 10^{-3}$ & $2 \times 10^{-3}$ & $2 \times 10^{-3}$ & \raggedright steady-state, $\dot{M}(t)$ from \citet{Hartmann1998} &  \\
		\noalign{\smallskip}\svhline\noalign{\smallskip}
		Disk Temperature$^\ddagger$ & {\raggedright viscous+irradiation heating}, 1\% dust-to-gas & {\raggedright viscous+irradiation heating}, varying dust-to-gas &  {\raggedright   viscous+irradiation heating}, 1\% dust-to-gas & \raggedright tables from 3D simulations \citep{Bitsch2015}, 0.5\% dust-to-gas, $\alpha_{\rm heat}$: $5.3\times 10^{-3}$ & \\
		\noalign{\smallskip}\hline\noalign{\smallskip}
		\raggedright Disk Photoevaporation &  \raggedright internal \citep{Clarke2001} \& external photoevaporation & \raggedright internal \citep{Clarke2001} \& external photoevaporation & \raggedright  internal \citep{Clarke2001} \& external photoevaporation & None, but decaying $\dot{M}$ & \\
		\noalign{\smallskip}\hline\noalign{\smallskip}
		\raggedright Migration (Type\,I / Transition / Type\,II) &  \raggedright Lindblad \& co-rotation / Crida / supr. disk-locked \citep{Dittkrist2014}  & \raggedright Lindblad \& co-rotation/ Crida / supr. disk-locked \citep{Dittkrist2014}  & \raggedright Lindblad, co-rotation, \& thermal ($\alpha_{\rm mig}$: $2 \times 10^{-4}$) / \citet{Kanagawa2018} &  Lindblad \& co-rotation ($\alpha_{\rm mig}$: $10^{-4}$) / \citet{Kanagawa2018} & \\
		\noalign{\smallskip}\svhline\noalign{\smallskip}
		\multicolumn{4}{l}{\textbf{Initial Conditions}}  & &\\
		\noalign{\smallskip}\hline\noalign{\smallskip}
		Initial embryo & \raggedright 0.01\,M$_{\oplus}$, log-uniform [$r_{\rm in}$,40\,AU] & \raggedright 0.01\,M$_{\oplus}$, log-uniform [0.1,50]\,AU & \raggedright 0.01\,M$_{\oplus}$, log-uniform [$r_{\rm in}$,$0.5 r_{\rm disk}$] & \raggedright Hill Regime, log-uniform [$r_{\rm in}$,$r_{\rm disk}$]  & \\
		\noalign{\smallskip}\hline\noalign{\smallskip}
		\raggedright Disk mass distribution & \raggedright \citet{Tychoniec2018} & \raggedright \citet{Tychoniec2018} & \raggedright \citet{Tychoniec2018} &  0.05\,M$_{\odot}$ for $t_{\rm disk} = 5$\,Myr&\\
		\noalign{\smallskip}\hline\noalign{\smallskip}
		\raggedright Disk Radius & \raggedright \citet{Andrews2010} & \raggedright \citet{Andrews2010} & \raggedright \citet{Andrews2010} & N/A & \\
		\noalign{\smallskip}\hline\noalign{\smallskip}
		\raggedright Solid mass content & \citet{santosisraelian2005} & \raggedright \citet{santosisraelian2005}, 90\% pebbles &  \citet{santosisraelian2005} & \raggedright Normal distribution, 0.23\,dex &\\
		\noalign{\smallskip}\hline\noalign{\smallskip}
	\end{tabular}
	\raggedright $^\dagger$:$f_{\rm opa} = 3\times 10^{-3}$ leads to minimum opacities on Earth-sized planets at 1\,AU of $\approx$\,0.005\,cm$^2$\,g$^{-1}$\\
	$^\ddagger$: All works use the tables from \citet{Bell1994} for the disk opacity
\end{table}
The initial conditions of three of the four models are almost identical and shown in Figure \ref{fig:ini_cond} (blue). The \citet{Tychoniec2018} masses with \citet{santosisraelian2005} metallicities have now been adopted in several planetary population synthesis works. However, the disk masses are therefore larger than what was considered in earlier generations of planetary population syntheses \citep[see][]{Mordasini2018}. \rev{Even more massive are the disk masses used in D23 ($\approx$\,150\,M$_{\oplus}$ on average with a 0.23 dex spread), similar to the orange histogram shown in Figure \ref{fig:ini_cond} suggested by \citet{Emsenhuber2023}. The argument for larger initial masses involves the unknown stellar mass of the observed sources which} implies an observational bias towards \rev{sub-Solar} stellar masses in observations. Overall, disk masses are for all models comparable and should not lead to large differences.

More significantly, model parameters differ. Table \ref{tab:models} lists {key parameters and initial conditions} for reference. \rev{Here, we briefly discuss the important differences skipping differing choices which are not expected to impact the results significantly.}
\begin{itemize}
		\item \rev{\textbf{Solid accretion.}} \rev{Between the two models using planetesimal accretion, the differences in the prescription for the accretion rates are minor with a slightly higher equilibrium eccentricity and inclinations assumed in \citet{Kimura2022a} compared to solving the time evolution thereof in NGPPS \citep{Fortier2013}. More importantly, the planetesimals are assumed to have a radius of 20\, km in \citet{Kimura2022a} while 300\,m radii were assumed in NGPPS. Due to the gas drag damping eccentricities and inclinations effectively for smaller objects, this results in significantly more favorable conditions for planetesimal accretion in NGPPS.} \\
		The \rev{two} pebble accretion models differ in their prescription for the accretion of pebbles. B20 follow \citet{Johansen2017} who in particular solve for the accretion rates by assuming drift-limited pebble sizes \citep{Lambrechts2014a}. In contrast, D23 use a constant Stokes number of 0.01 which is in the same ballpark as fragmentation limits for growth \citep[e.g.][]{Birnstiel2016}. Furthermore, the scale height of pebbles is a key parameter to determine. It depends on the pebble Stokes number, but also on the assumed vertical turbulence $H_{\rm p} = H \sqrt{\frac{\alpha_{\rm vert}}{{\rm St}+\alpha_{\rm vert}}}$. B20 assume the same value for $\alpha_{\rm vert}$ as used for their viscous gas evolution, while D23 assume a lower value as listed in Table \ref{tab:models}. For most conditions in the disk, these variations imply more efficient accretion and in the model of D23.\\
		\item \rev{\textbf{Treatment of orbital dynamics.}} Continuing in the order of Table \ref{tab:models}, the N-body effects considered differ between the \rev{two} planetesimal-based models. The analytic treatment of KI22 will be compared in more detail against the full N-body in NGPPS below. The pebble-based models do not consider N-body effects. This has two fundamental effects on the population level: (a) no growth via giant impacts and (b) no resonance-locking and thus no reduction of radial migration. Both effects have been shown to impact the distribution of low-mass planets fundamentally \citep{Alibert2013,Emsenhuber2020a}. Thus, when considering the low-mass planets in the pebble-based models, their masses should be seen as lower limits to planets in true systems.\\
		\item \rev{\textbf{Gas accretion. }}\rev{The treatment of gas accretion onto the growing planets is similar in the early stage in B20, NGPPS, and \citet{Kimura2022a} with only slight differences in choices of opacities (following \citealp{Mordasini2014} for the former two respectively \citet{Ormel2014} for the latter). For the comparison here, the water fraction, affecting mean density and opacities, was set to zero in the shown population of \citet{Kimura2022a} consistent with the assumption made in NGPPS and B20. All three works solve the structure equations for the envelope with the equation of state of \citet{Saumon1995}. In contrast, D23 use a prescription for gas accretion based no \citet{Piso2014} who used a more simple polytropic equation of state which leads to an underestimation of the critical core mass by a factor of two \citep{Piso2015}.}\\
		\item \rev{\textbf{Disk models.}} The protoplanetary disk models all share the same opacity prescription from \citet{Bell1994} which leads to opacity transitions at the same temperatures. Furthermore, the $\alpha$ values for viscous heating are equal in NGPPS, B20, and KI22. It is also only larger by a factor of 2.5 in the \citet{Bitsch2015} disks which are used by D23. The main temperature-related difference in the disk obtained from 3D calculations in \citet{Bitsch2015} is a shadowed region \rev{outside the water iceline implying lower temperatures and scale heights not included in the other models.}\\
		On the other hand, the models with the simpler temperature calculation are suitable to include non-steady-state disks which can be shaped into varying profiles by disk photoevaporation. They use the same simple photoevaporation prescriptions inspired by \citet{Clarke2001,alexanderpascucci2014} and \citet{Matsuyama2003} for internal and external photoevaporation, respectively. Furthermore, their initialization is more versatile with varying initial gas disk masses and radii (see \ref{tab:models}).
		The treatment of the disk and especially its temperature is significant for orbital migration rates of the planets. The four models \rev{use comparable migration prescriptions \citep{Paardekooper2010,Paardekooper2011,Crida2006,Dittkrist2014,Kanagawa2018} with the exception of thermal torques \citep{Masset2017} included only in \citet{Kimura2022a}. However, the viscosity used for migration $\alpha_{\rm mig}$ differ: Both KI22 and D23 use reduced $\alpha_{\rm mig}$} which suppresses the migration rate. This is similar to a migration suppression factor used in prior works \citep{Ida2008,Mordasini2009b} although more physically motivated \rev{by magnetized wind models.} By this argument, in consequence, the disk evolution equation \rev{and migration prescription} would then need to be \rev{modified (see the discussion in the Gas disk and Orbital dynamics and migration sections above)}.
	
\end{itemize}

\subsection{The $a-M$ distribution}\label{sect:amdistribution}
With these considerations on the model differences, a first look can be taken at the {resulting semi-major axis versus planetary mass distribution}. Such an analysis is classical in the field since \citet{Ida2004}. In Fig. \ref{fig:a_Mcomparison} the four varying populations are shown in addition to the Solar system planets \rev{and probability density estimates of the distribution of observed exoplanets. The comparison to observations is challenged by the observational bias. Nevertheless, by keeping the general trend of the bias in mind (planets are less likely to be observed if they are distant and small), we can discuss trends reproduced by the models. For a detailed quantitative analysis, the observational bias should be applied to the simulated planets to draw mock observations, which is exemplified in the Example comparison with observation: Kepler low-mass planets Section below.}

\rev{Starting with Hot Jupiters, it is important to remember that they are the most readily observable exoplanet category and in principle rare (only 0.5\%-1\% of stars host a Hot Jupiter). Given that the models simulate ~1000 systems, we would expect a handful of objects which is recovered by most models. To properly address Hot Jupiter statistics, more simulations would be required.}

\rev{Moving away from the star, bias corrected occurrences of giant planets generally increase from 0.1 to 3\,au before decreasing further out \citep{Fernandes2019,Fulton2021}. This is visible even in the direct statistics of observed exoplanets in the form of the second cold giant peak. The trend is relatively well matched by NGPPS, although the simulated peak is offset to slightly shorter orbits. The offset to shorter orbits is larger in B20 and also (tentatively due to few giants forming) in KI22. In contrast,the simulations from D23 show an accumulation of giants at larger distances. These differences are likely caused by the differences in migration parameters. A lower $\alpha_{\rm mig}$ value used in KI22 and D23 will lead to less significant migration but also no emergence of a migration trap due to a weakening of the co-rotation torque explaining the lack of a pileup in KI22. In D23, the combination with the higher efficiency of pebble accretion compared to planetesimal accretion leads to more distant giant planet formation.} Overall, the number of giants is strongly reduced in B20 and enhanced in D23 over the two planetesimal-based models. This was also discussed in \citep{Drazkowska2023} \rev{and can be attributed to be caused by the parameter differences between B20 and D23 and the lower efficiency of planetesimal accretion at separations > 1\,au.}

Fast migration is also detrimental to growth in the type I migration regime around 10\,$\mearth$. For high enough $\alpha_{\rm mig}$ values, this effect limits growth in NGPPS and B20 and to a lesser degree also in D23. While the migration of 10\,$\mearth$ mass sub-Neptunes seems to be in agreement with the low-mass peak in the probability density estimate of the observed planets (gray contours), this is heavily influenced by the reducing transit probability with distance which is the most common method to detect low-mass exoplanets. A comparison to Kepler data reveals (see discussion below or \citealp{Burn2024}) that NGPPS and thus likely also B20 and D23 over-estimate the number of close-in low-mass planets. This motivates a further reduction or sporadic interruption \citep{Coleman2016a} of (at least type I) migration.

\begin{figure}[h!]
    \centering
    \includegraphics[width=0.99\textwidth]{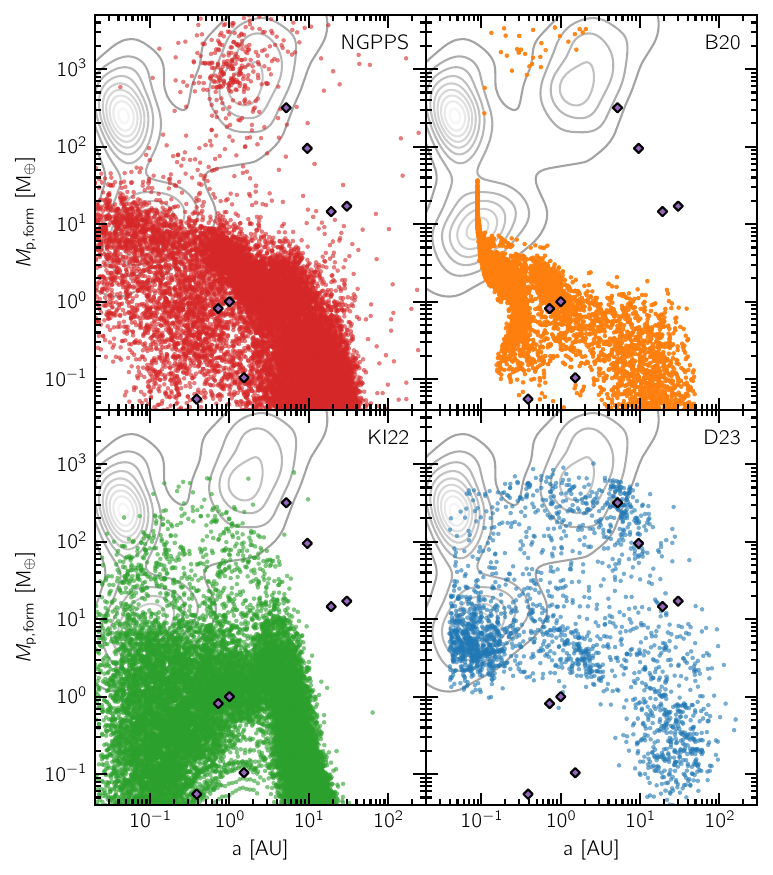}
   \caption{Synthetic mass-distance diagrams for different theoretical models. \rev{From top left to bottom right the panes show results obtained using the Bern Model \citep{Emsenhuber2020a,Emsenhuber2020b} (NGPPS)}; from \citet{Brugger2018,Brugger2020} (B20); \citet{Kimura2022a} (KI22); and \citet{Bitsch2015a,Drazkowska2023} (D23). The mass and location are taken after the formation stage, that is, not including long-term mass loss processes and tidal migration. For orientation, \rev{a 2D Gaussian kernel density estimate with 0.18 dex bandwidth (see also Fig. \ref{fig:mkdecomparison}) of the observed exoplanet distribution (same as Fig. \ref{CMaMEpoch2017}) and} the Solar system planets are shown. Note the unequal number of planets in the four model results.}
   \label{fig:a_Mcomparison}
\end{figure}

The width of the gap in planetary mass between super-Earths and sub-Neptunes at $\sim$\,10\,M$_{\oplus}$ and the giant planets ($M_p > 100\,$M$_\oplus$, the so-called runaway gas accretion desert or planetary mass desert \citealp{Ida2004}) -- \rev{whose observational confirmation is} disputed \citep{Mayor2011,Bennett2021} -- differs considerably. \rev{Since it is expected to be caused by rapid gas accretion, insights gained from the comparison to observation can constrain maximum gas accretion rates and envelope cooling processes.} In NGPPS and B20, the feature bridges more than an order of magnitude in mass. Despite the similar gas accretion model (see Table \ref{tab:models}), the giant planets in B20 are \rev{on average more massive than in NGPPS. This is related to earlier emergence of the planets in B20 where both the local gas reservoir is larger but migration rates are also faster leading to the emergence of a massive giant whenever a growing planet jumps the higher migration "hurdle". The more narrow and populated planetary desert in D23 and K22 is likely also related to this migration barrier but there are also differences in the maximum gas accretion rate prescription which should be investigated further.}

\rev{To summarize, the models do apparently not fit the observed distribution of exoplanets, which is however to a large degree due to the observational bias. When considering this, the D23 and NGPPS simulations show features in agreement with observations such as an increasing giant planet occurrence with distance, a matching high number of close-in small planets (see also discussion below), and a planetary desert bridging the lack of exoplanets between 10 and 100\,$\mearth$.}

\begin{figure}
	\centering
	\includegraphics[width=.75\linewidth]{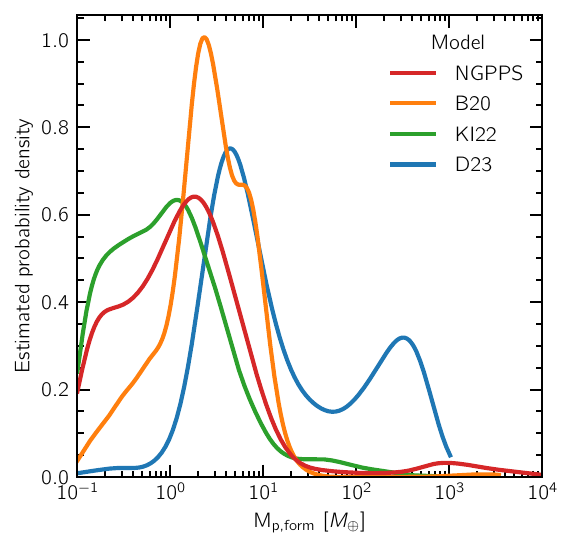}
	\caption{Kernel density estimates of four synthetic planetary populations. \rev{A probability density function is estimated using the sum over smooth functions for each predicted planetary mass, where the smoothing function is a Gaussian distribution centered at the planet mass with a dispersion (bandwidth) chosen to be 0.2 dex \citep[see \texttt{scipy.stats.gaussian\_kde}][]{scipy}.} Planets below 0.1\,M$_{\oplus}$ were removed before the smoothing. }
	\label{fig:mkdecomparison}
\end{figure}

\subsection{The planetary mass function}

The {resulting planetary mass distribution} of the four planetary population synthesis models compared here is shown in Fig. \ref{fig:mkdecomparison}. As mentioned above, the giant planet population is more numerous in the pebble accretion model of D23 (blue) \rev{and the differences in the planetary desert are recovered in this one dimensional view}.

Focusing on the lower-mass population, new insights can be gained. The B20 population has a slightly bi-modal distribution which is an imprint of migrated planets from colder regions with larger pebble isolation masses and the in-situ inner-disk planets. These differences were also recovered by \citet{Venturini2020} and could be related to the radius valley \citep[see also][]{Izidoro2022} if there is a corresponding mass-valley in nature, which is so-far not accessible statistically from radial velocity surveys.

The mass functions of the lower-mass planets in NGPPS and KI22 are uni-modal, broader and peak below the one of the pebble accretion models. This was already \rev{found} in \citep{Brugger2020}. \rev{Since larger planetesimals are assumed to exist in KI22, their eccentricities and inclinations are less efficiently damped by gas drag and they are less likely to collide with each other or a growing protoplanet (see Solid accretion Section). Thus}, their mass function lies below the one obtained in NGPPS. However, \rev{the two} shapes are similar apart from the typical giant planet mass \rev{influenced by a combination of solid accretion rates, migration, and gas accretion}.

\subsection{Diversity of planetary system architectures}\label{sect:diversityofplanetarysystemarchitectures}

\begin{figure}
	\centering
	\includegraphics[width=1.\linewidth]{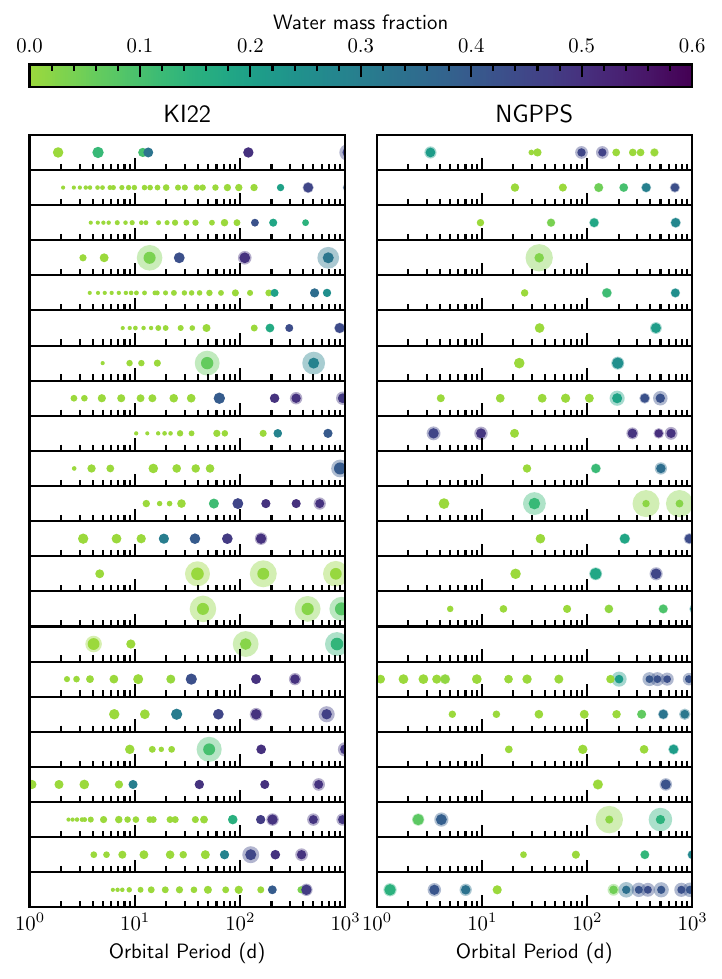}
	\caption[Comparison of synthetic system architectures]{Comparison of the architecture of 25 planetary systems from the theoretical models by \citet{Kimura2022a} and NGPPS. The water fraction by mass in the planets is color-coded. The transparent circles are proportional to the total radii, the full circles to the core radii of the planets. Compositional inside-out ordering from rocky to wet planets is paramount but can more easily be broken in models with full N-body interactions. A difference between model results exists in the typical mass of giants (NGPPS giants are more heavy) and the varying core radius (core contraction) as discussed in the text. }
	\label{fig:systems_ki_ng}
\end{figure}

For the two population synthesis models with multiple embryos per disk, a brief overview of the various {resulting planetary system architectures} can be given. Since in both works, the long-term evolution is included, the comparison is done here after evolution for 5\,Gyr. This is mainly relevant for the radii of -- and presence of envelopes on -- intermediate mass planets.

The analysis of planetary system architectures can have a considerable complexity and modern ways and criteria to distinguish planetary systems were recently developed and reviewed \citep{Mishra2023,Mishra2023a,Emsenhuber2023a,Weiss2023}. For the sake of brevity, the first 25 tabulated systems of the two models are compared visually in Fig. \ref{fig:systems_ki_ng}. In general, a trend of smaller, more water poor planets populating the inner systems can be identified. Although this overall trend is present, when considering only planets on short orbits, as done by observations, many systems in both models show a 'peas-in-a-pod' \citep{Weiss2018} architecture with similar sizes and regular spacings. \citet{Mishra2021} scrutinized the architectures of the NGPPS population and contrasted them to the Kepler mission to recover this conclusion.

By visual comparison, it can be seen that the low-mass inner systems of KI22 often contain more planets with narrower orbital spacing. This could be a limitation of the analytic approach for N-body interactions. In NGPPS, the planets in those systems grew in-situ by planetesimal accretion before transitioning to a giant impact stage where they grew by a factor of a few. Mutual eccentricity excitation can be invoked to construct a mass scale which characterizes this stage \citep{Emsenhuber2023a}.

Further, water-rich planets do not switch positions with dry super-Earths in KI22 while this is a relatively common occurrence in NGPPS. The method to compute the next close-encounter in KI22 follows \citet{Ida2010}. There, several assumptions are made, which include to excite the eccentricities of a pair of planets with the shortest orbit-crossing time first, then identify groups of embryos which now interact (and may collide) and assign increased eccentricities to the whole group of planets with average values for all but the most massive planet in the group. As \citet{Ida2010} mention, this leads to a diffusion of embryos but cannot lead to changes in their ordering. For now, analysis of the ordering and compositional gradients thus requires conducting full N-body calculations. Another difference influencing this comparison is that of differing migration rates in the two models which also has an influence on how common migration of water-rich planets to the inner system is.

Lastly, there is an interesting difference in the core radii of giant planets. This is a rather minor detail but prominent in Fig. \ref{fig:systems_ki_ng}. Core contraction is included in NGPPS but not in KI22. Furthermore, the giants in NGPPS are on average more massive (as mentioned in the discussion of the a-M diagram), thus their cores get contracted more significantly.

\subsection{Example comparison with observations: Kepler low-mass planets}\label{sect:compobsdist}
The discussion here now turns to exemplifying the key aspect in planetary population synthesis, which is the {comparison of model predictions to observations}. As an example, the NGPPS population \rev{with updated radii obtained by using more realistic treatment and distribution of water and photoevaporation in hot exoplanet atmospheres \citep{Burn2024}} is contrasted against data from the Kepler space telescope. This exercise will join the studies of \citet{Emsenhuber2020b}, \citet{Burn2021}, \citet{Mishra2021}, \citet{Schlecker2022}, \citet{Desgrange2023}, \citet{Eberhardt2023}, and \citet{Burn2024} who conducted comparisons of the NGPPS simulations to selected observations. More works are in preparation.

\rev{The comparison of }planetary radii in the low-mass regime is a more challenging and modern aspect of exoplanetary sciences.\rev{The parameter space is statistically accessible} thanks to space-based transit searches; the state-of-the art is provided by the Kepler mission \rev{\citep{Borucki2010}}. To allow for comparison of theory with observations, it is required to carefully assess the observational bias. For a mission such as Kepler, which provides detailed statistics, it would be \rev{sub-optimal} to use too simplistic approaches since those can lead to wrong conclusions given the constraining power of the observations. \rev{Simple estimates of the transit ($p_{\rm tr} \simeq 0.9 R_{\star}/a$) and detection probability $p_{\rm det}$ \citep[e.g. Eqs 12-13 in][]{Petigura2018} can however be useful for quick analysis. Since it is often too involved to use the full Kepler stellar catalog to determine variabilities, we estimate the average (over stars) detection probability for the Kepler mission based on the data shown in \citet{Petigura2018} as
\begin{equation}
p_{\rm det} \approx \Phi\left( R/\rearth, \mu = 1.3866 \frac{P}{100\,d}, \sigma = 0.145 \right)\,,
\end{equation}
with $\Phi$ being the cumulative distribution function of the normal distribution with mean $\mu$ and standard deviation $\sigma$.}

\rev{Instead of relying on simple estimates, we use here} the KOBE code \citep{Mishra2021}, which can be obtained in open access \citep{KOBEURL}. \rev{It }is used to obtain mock observations with Kepler from synthetic populations. The package attributes random orientations to the synthetic planetary systems and uses the thresholds of the Kepler DR25 pipeline as described in \citet{Mishra2021}. 

\begin{figure}
	\centering
	\includegraphics[width=\linewidth]{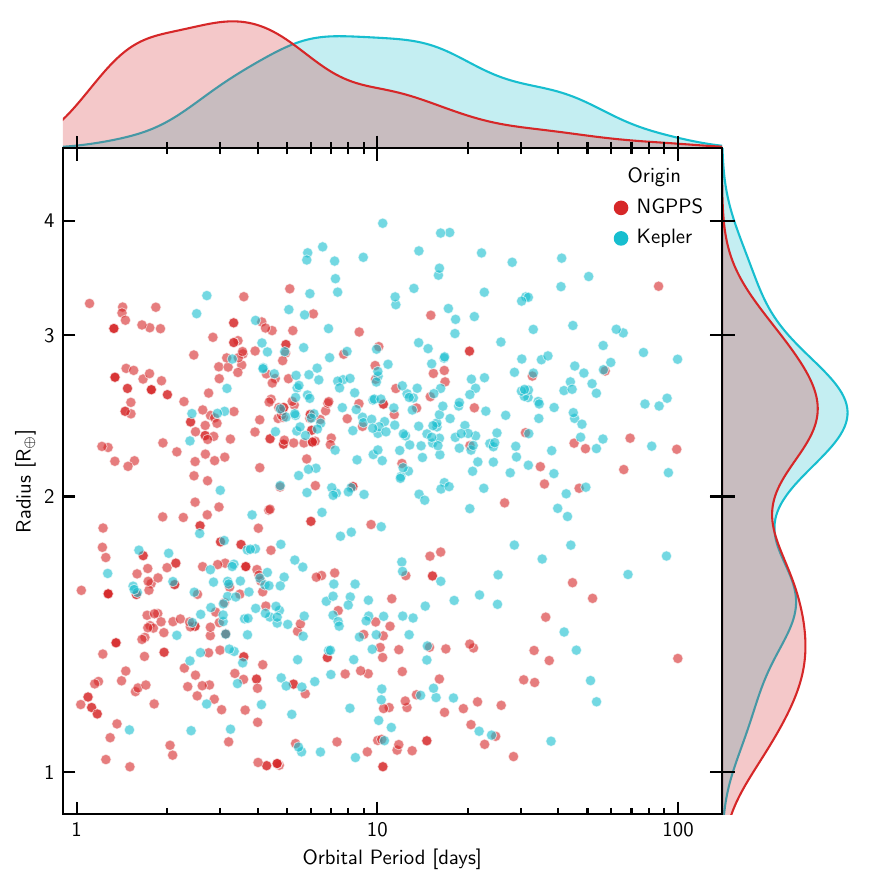}
	\includegraphics[width=0.6\linewidth]{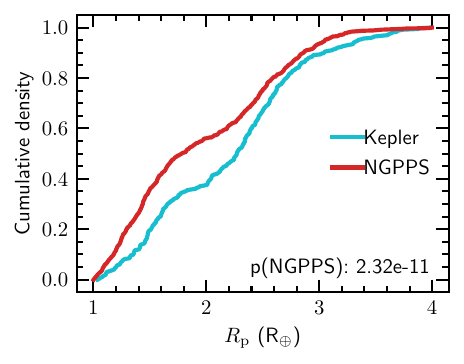}
	\caption{Comparison of synthetic and observed radii and orbital periods for small planets. The observational data is originally from the Kepler spacecraft but vetted, selected, and improved as described in \citet{Ho2023}. They restricted the sample to $1<P<100$\,days and $1<R<4$\,$R_{\oplus}$. For the synthetic data, the observational bias from the Kepler spacecraft was applied. Finally, the synthetic data is cut to the same regime and the same number of planets as observed is randomly drawn. The stellar radius assumed equals that of a Solar mass star after 5 Gyr of evolution following \citet{Baraffe2015}. The top panel includes one draw of the same number of planets and a kernel density estimate to the distributions with a fixed bandwidth of 0.3 dex. In the bottom panel, the cumulative distributions of the full data (including no draws for the synthetic distributions) are shown. The $p$-value of a statistical Kolmogorov-Smirnov test is indicated at the bottom right.}
	\label{fig:kepler_comparison}
\end{figure}

For the observational data, the table provided by \citet{Ho2023} was used, which builds upon the California Kepler Survey \citep{Petigura2017,Fulton2018} -- the program that originally discovered the radius valley in observations \citep{Fulton2017}. The observed radius valley can be clearly seen in Fig. \ref{fig:kepler_comparison} as well as a general increase of planet number up to orbital periods of $\sim 7$\,days \citep[see also][]{Mulders2015b}. Another known trend is the longer orbital period of the innermost planets with radii above the valley (i.e. the sub-Neptunes) compared to the innermost super Earths (the planets below the valley), which can be understood as the lower end of the photoevaporative desert \citep{jinmordasini2018,Owen2018}.

\rev{Also the simulated planetary distribution} shows a prominent {radius valley} at the location where it is observed. \rev{As found by \citet{Venturini2020,Venturini2024,Burn2024}, }the valley \rev{can} emerge as the separation from dry super Earths which formed almost in-situ to water-rich, migrated sub Neptunes above the radius valley. 

Other works used interior models to ascertain that such an interpretation of the radius valley is possible \citep{Mousis2020,Zeng2019,Zeng2021}. \rev{This provides a formation-theory-motivated alternative to the interpretation of the radius valley being carved by photoevaporative H/He loss as the only process \citep{Jin2014,Jin2018,Owen2013,Lopez2013,Rogers2021,Mordasini2020,Owen2017,Owen2024}}.

From Fig. \ref{fig:kepler_comparison}, we see that the relative number of sub-Neptunes is lower than observed compared to the super Earths with the opposite trend. \rev{An} over-production of small planets, could indicate too efficient growth by planetesimal accretion in the inner system. Alternatively, the number of planets which are stripped from their primordial envelopes could be too large, since this alternative pathway to become a rocky super Earth is also common in the population.

Differences like these can also be quantified. A commonly adopted statistical test to compare models to observations is the Kolmogorov-Smirnov test \citep{Raymond2009,Mordasini2009b,Alibert2011}. Here, we use the implementation from the \texttt{scipy} python package (\texttt{scipy.stats.ks\_2samp}, \citealp{scipy}) to exemplify such a comparison in planetary radii. The bottom panel of Fig. \ref{fig:kepler_comparison} shows the resulting comparison with the statistical conclusion that the NGPPS distribution of planetary radii differ in a significant way. To quantify this, it is insightful to use the $p$ values of this test as a measure to how close the models are to a statistically insignificant difference, that is, a reproduction of observations within statistical limits. In principle, observational measurement errors should be accounted for. Here, the difference between the synthetic and observed distribution is large and the observational errors are small, therefore more detailed calculations are not required. For seemingly matching distributions, bootstrapping of the observational or synthetic sample can be done to obtain a distribution of $p$-values. Furthermore, it is good practice to include various statistical tests such as the Cucconi or the Wilcoxon test in addition to the Kolmogorov-Smirnov one.

The too short orbital periods of sub-Neptunes in NGPPS seen here is also seen in \citet{Schlecker2022} where RV planets are located closer to their host stars than observed for high stellar masses. Interestingly, around late M dwarfs, planets seem to be located closer to their host stars. This might be due to the higher sensitivity probing into the rocky planet, low-mass regime.

While the conclusion can be drawn that the population synthesis needs to be improved to match better the observations from Kepler, it is noteworthy that for some regimes, that is, at short orbital periods, the NGPPS results discussed in \citet{Burn2024} differ insignificantly from the observed distribution. This is a promising sign that planetary population synthesis can at least partially reproduce observations and is not trailing far behind advances in observations.

\section{Outlook}
\subsection{Viscosity prescriptions}
In the previous sections, a few ongoing and required developments were already hinted at. One important factor in the analysis of the different planetary population synthesis models is the choice of {$\alpha$ parameters} which we briefly discuss here. In a viscous disk, $\alpha$ responsible for the angular momentum transport (\citealp{Shakura1973}, in this discussion called $\alpha_{\rm S}$) is given by the Reynolds and Maxwell (if there are magnetic fields involved) stress of some turbulence \citep[e.g.][]{Armitage2019,Lesur2023}. However, in addition to the angular momentum transport, other $\alpha$ parameters are required to calculate the collision velocity between dust $\alpha_{\rm small}$ \citep{Ormel2007}, for the radial diffusivity of the gas and the dust $\alpha_{\rm d,r}$ \citep[e.g.][]{Birnstiel2016}, to equilibrate settling of dust against turbulent vertical stirring of dust $\alpha_{\rm d,z}$ or $\delta_z$ \citep{Youdin2007}, to describe the viscous flow of material in the midplane region around planets $\alpha_{\rm mig}$ \citep[related to migration, discussed above, see e.g.][]{Paardekooper2023}, and to prescribe viscous heating in the midplane $\alpha_{\rm heat}$. In a realistic disk, these six parameters could all be distinct from each other as well as vary with radial location \citep{Lesur2023}. For example, turbulence does not necessarily follow a Kolmogorov kinetic energy spectrum \citep{Gong2020}, which is the energy in the different turbulent modes of various length-scales. This would imply a different $\alpha$ for small-scale processes, such as dust collisions. Other modes of turbulence are highly anisotropic, such as the one driven by the vertical shear instability (VSI). This instability produces extended vertical modes but modes with much shorter extent in the radial \citep{Dullemond2022,Lesur2023}.

\rev{However, treating all these processes as free parameter is not a useful approach as the disk model would then be severely under-constrained. Thus, it is required to find relations between the different turbulent strenghts. For the VSI, the vertical $\delta_z$ } could be modeled a factor 100 \rev{larger than} $\alpha_{\rm d,r}$ and $\alpha_{\rm S}$ \citep{Dullemond2022}. It is probably also not efficient in heating the midplane of the disk implying an even lower $\alpha_{\rm heat}$ (H. Klahr, D. Melon Fuksman, private communication). For the VSI, a prescription is now available for its strength as a function of disk opacity and temperature gradient \citep{Manger2021}, which could be included in future population synthesis works. Similarly, the magnetorotational instability can be described in more detail to obtain radial dependencies \citep{Delage2022,Delage2023}. For further discussion, see also the Chapter by Klahr in this volume.

As discussed above, in a {wind-driven disk}, the disk does not follow a diffusion equation anymore, which should be modeled correspondingly. The description of migration in these disks is the subject of ongoing investigations \citep{Paardekooper2023} and we showed above how critical it is for the outcome of population synthesis models. For now, one may think that that when modeling truly viscous disks, $\alpha_{\rm mig}$ should be similar to $\alpha_{\rm S}$, since the vertically integrated column of mass contributes to the torques. For primarily wind-driven disks \citep{Alessi2022,Weder2023}, a remnant $\alpha_{\rm S}$ should be included which could then also be used for viscous migration prescriptions until the impact of the winds have been fully characterized.

For population synthesis, the development of a physically-informed viscosity prescription is important. The differences found in the comparison between population synthesis models discussed above can be attributed to a large degree to the difference in non-dimensional viscosity parameters. Also, the temperature structure -- strongly impacted by $\alpha_{\rm heat}$ in the inner few au -- is especially important to determine gas-scale-height-dependent processes such as migration and the pebble isolation mass \citep{fouchetalibert2012}.

\subsection{Solid accretion}
Both planetesimal accretion and pebble accretion face theoretical challenges}. In both cases, this has to do with the advances in planetesimal formation modeling. Clumping of dust by gravitational forces requires a certain threshold density of dust before the collapse can occur. Therefore, clump sizes cannot be arbitrarily small. As discussed above, prescriptions of the dust clumps are now available and have been used to model planetary formation. However, follow-up simulations show that the typical compact planetesimal might be of order hundred kilometers in size which is in agreement with asteroids identified to be primordial \citep{Delbo2019,Polak2023}. This size poses a {challenge to both planetesimal and pebble accretion} scenarios. It is too small to efficiently grow by pebble accretion, but too big to be subject to efficient aerodynamic drag. The lack of drag means that planetesimal inclination and eccentricity can grow, which reduces mutual collision rates. Especially in the outer system, this implies in smooth disks that no larger embryos can form. \citet{Voelkel2021,Voelkel2021a} investigated this scenario and could not form distant giant planets. This is in disagreement with the typical initialization of population synthesis models with large embryos. Fragmentation of planetesimals to intermediate sized fragments could help to resolve the issue to some degree \citep{Guilera2014,Kaufmann2023}. Nevertheless, future models will likely transition to structured disks (since they are also observed) to address these issues but this creates a large set of free parameters, which is detrimental to the population synthesis approach. Physical mechanisms for substructures are required for a more concise picture, such as icelines \citep{Drazkowska2017}, turbulence transitions like the dead zone outer edge \citep{Pinilla2016,Delage2023}, or preferential infall locations \citep{Kuznetsova2022}.

\rev{Another path forward can be to use the forward-modeling approach with planetary population synthesis and compare their envelope compositions. Even more consistent, in the spirit of taking a step towards the observations, is to generate mock transmission spectra using synthetic planets, which will be an important task for planetary population syntheses to make best use of observational data.}

\section{{Summary and conclusions}}
Since the review of \citet{Mordasini2018}, numerous developments have been made on the observational and theoretical side \rev{of exoplanet demographic studies. Here, we summarize the approach of planetary population synthesis which aims at reproducing the statistical distribution of exoplanet observables -- such as the mass, radius and distance to the star. This requires a global model of planet formation which can form the final, evolved planets from observationally constrained initial dust and gas disks. Here, we outlined the main ingredients of these models and the available observational data on both disks and exoplanets. To ascertain an unbiased confrontation of theory and observation, the observational bias needs to be accounted for.}

\rev{The approach has led to the successful prediction or confirmation of several demographic features. An established result is the dependency of giant planet occurrence on the stellar metallicity as a proxy for the available solid mass for growth \citep{gonzalez1997,Santos2001,santosisraelian2004,Ida2005,Mordasini2009b}. A stellar-mass dependent threshold metallicity is predicted also for lower planetary masses \citep{Burn2021}. Moreover, the radius valley \citep{Fulton2017} shaped by photoevaporative mass loss was predicted using a population synthesis approach \citep{Jin2014}. In recent years, the approach was used to contrast different planet formation models, relying on pebble or planetesimal accretion, to find a distinct impact on the planetary population \citep{Bitsch2017,Brugger2020}. A major refinement on the initial conditions could be made owing to comprehensive surveys of -- especially young -- protoplanetary disks \citep{Tychoniec2018,Tychoniec2020,Tobin2020} leading to a higher predictive power of the approach. Moreover, this led to the inclusion of consistent dust evolution in thus truly global models tracing growth from grain to planet \citep{Alessi2020,Voelkel2020,Coleman2021,Schneider2021,Drazkowska2021}.}

By comparing four different planetary population syntheses \rev{utilizing} different models in detail \rev{to each other and to observations}, we naturally obtain the pathways for {future works}: There is the need to address the emergence of embryos of sufficient size to trigger pebble accretion. Furthermore, in the new disk paradigm driven by magnetic winds, the orbital migration has to be better characterized and new theoretical work on this front needs to be included in global models. In those aspects, the reviewed models differ significantly and the groups optimized parameters in different ways to reproduce observed planets. \rev{Although the gloabl modeles were developed}, dust growth and drift has not been modeled comprehensively for planetary population synthesis and should be applied to full population studies in the future.

An opportunity to advance our understanding of planet formation emerges with the improved constraints on disk and planets and their compositions from observations with JWST. The models will have to improve their descriptions of material composition to allow for detailed and comprehensive comparison \rev{while the field overall addresses the initial disk composition and how to link atmospheric to bulk abundances}.

\vspace{1cm}

\small{Acknowledgements: We thank T. Kimura and B. Bitsch for sharing with us their population data for this review which enabled a detailed comparison. Further, we thank Hubert Klahr for fruitful discussion. R.B. acknowledges financial support from the German Excellence Strategy via the Heidelberg Cluster of Excellence (EXC 2181 - 390900948) ``STRUCTURES'' under Exploratory Project 8.4. C.M. acknowledges the support from the Swiss National Science Foundation under grant BSSGI0$\_$155816 ``PlanetsInTime''. Parts of this work have been carried out within the framework of the NCCR PlanetS supported by the Swiss National Science Foundation under grant 51NF40$\_$205606.}


\bibliographystyle{aasjournal}
\bibliography{liball.bib,library_betterbib_zotero.bib,cmadd.bib}{}

\end{document}